\documentclass{pasj00}

\usepackage{epsfig}
\usepackage{subfigure}
\usepackage{graphicx}
\usepackage{rotating}
\usepackage{natbib}
\usepackage{booktabs}

\usepackage{threeparttable}

\Received{2014 May 30}
\Accepted{2014 ??}
\begin{document}
\title{Determination of Three-dimensional Spin--orbit Angle\\
 with Joint Analysis of Asteroseismology, Transit Lightcurve, \\
    and the Rossiter--McLaughlin Effect: Cases of HAT-P-7 and Kepler-25}
\author{Othman \textsc{Benomar}\altaffilmark{1},
Kento \textsc{Masuda}\altaffilmark{2},
Hiromoto \textsc{Shibahashi}\altaffilmark{1},
and Yasushi \textsc{Suto}\altaffilmark{2,3}}
\altaffiltext{1}{Department of Astronomy, The University of Tokyo,
School of Science, Tokyo 113-0033}
\altaffiltext{2}{Department of Physics, The University of Tokyo,
School of Science, Tokyo 113-0033}
\altaffiltext{3}{Research Center for the Early Universe, 
School of Science, The University of Tokyo, Tokyo 113-0033}

\email{othman.benomar@astron.s.u-tokyo.ac.jp}
 

\maketitle

\begin{abstract}
We develop a detailed methodology of determining three-dimensionally the
angle between the stellar spin and the planetary orbit axis vectors,
$\psi$, for transiting planetary systems.  The determination of $\psi$
requires the independent estimates of the inclination angles of the
stellar spin axis and of the planetary orbital axis with respect to the
line-of-sight, $i_\star$ and $i_{\rm orb}$, and the projection of the
spin--orbit angle onto the plane of the sky, $\lambda$.  These are
mainly derived from asteroseismology, transit lightcurve and the
Rossiter-McLaughlin effect, respectively. The detailed joint analysis of
those three datasets enables an accurate and precise determination of
the numerous parameters characterizing the planetary system, in addition
to $\psi$.

We demonstrate the power of the joint analysis for the two specific
systems, HAT-P-7 and Kepler-25. HAT-P-7b is the first exoplanet suspected to be a retrograde (or polar) planet because of the
significant misalignment $\lambda \approx 180^\circ$.  Our joint
analysis indicates $i_\star \approx \timeform{30D}$ and $\psi
\approx 120^\circ$, suggesting that the planetary orbit is closer
to polar rather than retrograde.  Kepler-25 is one of the few
multi-transiting planetary systems with measured $\lambda$, and hosts
two short-period transiting planets and one outer non-transiting planet.
The projected spin--orbit angle of the larger transiting planet,
Kepler-25c, has been measured to be $\lambda \approx 0^\circ$, implying
that the system is well-aligned. With the help of the tight constraint
from asteroseismology, however, we obtain
$i_\star=\timeform{65D.4}^{+\timeform{10D.6}}_{-\timeform{6D.4}}$ and
$\psi=\timeform{26D.9}^{+\timeform{7D.0}}_{-\timeform{9D.2}}$, and
thus find that the system is actually mildly misaligned.  This is the
first detection of the spin--orbit misalignment for the multiple
planetary system with a main-sequence host star, and points to
mechanisms that tilt a stellar spin axis relative to its protoplanetary
disk.
\end{abstract}

\section{Introduction 
\label{sec:intro}}

The spin--orbit angle between the stellar spin and the planetary orbital
axes, $\psi$, is supposed to be a unique observational probe of the
origin and evolution of planetary systems.  The existence of
Jupiter-like planets with orbital periods less than a week strongly
indicates that the inward migration of those planets is a basic
ingredient of successful theories of planet formation and evolution. A
fairly popular scenario of the migration is based on the planet--disk
interaction, 
in which planets are supposed to be on circular orbits whose orbital axes are parallel to the
stellar spin axis.  On the other hand, scenarios such as planet--planet
scattering 
or the Kozai mechanism 
predict a broad range of eccentric and
oblique orbits.  Thus the precise determination of $\psi$ and its
statistical distribution put a tight constraint on the viable migration
models \citep{Queloz2000, Winn2005}.

While the measurement of
$\psi$ is not easy, its projection onto the plane of the sky, $\lambda$,
has already been measured for more than 70 transiting planetary systems
via the Rossiter--McLaughlin (RM) effect \citep{2011exop.book...55W}, and is now
established as one of the most basic parameters that characterize
transiting planetary systems.
 
The RM effect was originally proposed to determine the projected
spin--orbit angle of eclipsing binary star systems \citep{R1924,M1924}.
\citet{Queloz2000} successfully applied the technique for the first
discovered transiting exoplanetary system, HD\,209458, and obtained
$\lambda = \pm \timeform{3D .9} {+18^\circ \atop -21^\circ}$.
In the quest for
improving the precision and accuracy, \citet{Ohta2005} presented an
analytic formula to describe the RM effect and studied in detail the
error budget and possible degeneracy among different parameters. This
allowed \citet{Winn2005} to revisit HD\,209458 with updated photometric
and spectroscopic data, and to obtain 
$\lambda=-\timeform{4^D.4} \pm \timeform{1D.4}$,
improving the precision of the previous measurement by an order of
magnitude.

In doing so, \citet{Winn2005} pointed out that the analytic
approximation adopted by \citet{Ohta2005} leads to typically 10 percent
error in the predicted velocity anomaly amplitude, while the estimated
$\lambda$ is fairly reliable.  This motivated \citet{Hirano2010} and
\citet{Hirano2011} to take into account stellar rotation,
macroturbulence, and thermal/pressure/instrumental broadenings in
modeling the stellar absorption line profiles. Those authors derived an
analytic formula for the velocity anomaly of the RM effect by maximizing
the cross-correlation function between the in-transit spectrum and the
stellar template spectrum. As a result, their analytic formulae reproduce
mock simulations within $\sim 0.5$ percent, enabling
the accurate and efficient multi-dimensional fit of parameters
characterizing the star and planet(s) of an individual system.

More importantly, \citet{Winn2005} clearly demonstrated the potential of
the RM effect to put strong quantitative constraints on the existing
and/or future planetary formation scenarios.  Indeed, when HD\,209458
was the only known transiting planetary system, \citet{Ohta2005}
discussed that {\it ``Although unlikely, we may even speculate that a
future RM observation may discover an extrasolar planetary system in
which the stellar spin and the planetary orbital axes are anti-parallel
or orthogonal.  Then it would have a great impact on the planetary
formation scenario, \ldots ''}. In reality, however, they were too
conservative. Among the 70 transiting planetary systems observed with
the RM effect, more than 30 systems exhibit significant misalignment
with $|\lambda|>\timeform{22D.5}$ [\,see, e.g., figure 7 of
\citet{Xue2014}\,].  This unexpected diversity of the spin--orbit angle
is not yet properly understood by the existing theories, and remains an
interesting challenge \citep[see,
][]{Fabrycky2007,Nagasawa2008,Winn2010b,Nagasawa2011,Hirano2012,Hirano2012b,
Lai2012,Albrecht2013,Masuda2013,Xue2014}.

\begin{figure}
	\centering
	\includegraphics[width=8.5cm,clip]{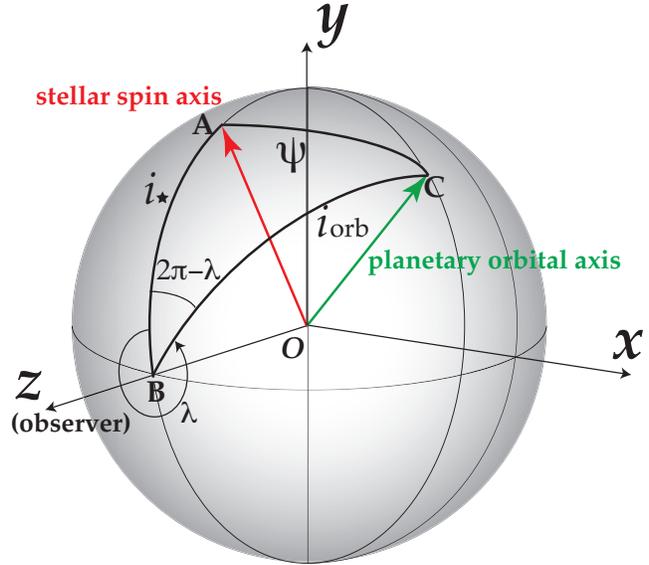}
	\caption{Schematic illustration of geometric configuration of
	star--planet systems.  We choose a coordinate system centered on
	the star, where the $XY$-plane is in the plane of the sky and
	$+Z$-axis points towards the observer.  The $+Y$-axis is chosen
	along the sky-projected stellar spin and the $X$-axis is
	perpendicular to both $Y$- and $Z$-axes, forming a right-handed
	triad.  Red and green arrows indicate on a unit
	sphere, the angular momentum vectors of the stellar spin and
	the planetary orbital motion, respectively.  The stellar and
	orbital inclinations, $i_\star$ and $i_{\rm orb}$, are measured
	from the $+Z$-axis and in the range of $[\,0^\circ, 180^\circ]$.
	The planetary orbital axis projected onto the sky plane is
	specified by the projected spin--orbit angle, $\lambda$, which
	is measured from the $+Y$-axis and in the range of $[0^\circ,
	360^\circ]$.  Note that $\lambda$ is measured in the
	direction specified by the arrow.  The angle AOC
	between the stellar spin and the planetary orbit axis vectors,
	$\psi$, is derived from the law of cosines for the spherical
	triangles ABC, as given by Equation (\ref{psi_3d}).}
	\label{angles}
\end{figure}

It should be noted, however, that $\lambda$ differs from the true
spin--orbit angle $\psi$ due to the projection on the sky.  In addition
to $\lambda$, $\psi$ also depends on the orbital inclination $i_{\rm
orb}$ and the obliquity of the stellar spin-axis $i_\star$:
\begin{equation}
\cos\psi = \cos i_\star \cos i_{\rm orb} 
+ \sin i_\star \sin i_{\rm orb} \cos\lambda ,
\label{psi_3d}
\end{equation}
as illustrated in Figure \ref{angles}.

The main purpose of this paper is to establish a methodology to
determine $\psi$, instead of $\lambda$, through the
joint analysis of asteroseismology, transit lightcurve, and the RM
effect, and to present specific results for a couple of interesting
transiting planetary systems, HAT-P-7\footnote{ We would like to
emphasize the efforts made by Lund M. N. and his collaborators for their
work on HAT-P-7. This system turned out to be studied simultaneously and
independently by our respective teams.} (KIC\,10666592) and Kepler-25
(KIC\,4349452).

In the case of transiting planetary systems, $i_{\rm orb}$ can be
estimated from the transit lightcurve, and in any case is close to
$90^\circ$.  Given the projected angle $\lambda$ measured from the RM
effect, the major uncertainty for $\psi$ comes from the unknown
stellar inclination $i_\star$.

There are two major and complementary approaches to estimate
$i_\star$, and hence $\psi$. One is to determine the line-of-sight
rotational velocity of the star, $v \sin i_\star$, either from the width
of absorption lines or from the RM effect.  The observed $v\sin i_\star$
is then compared with an independent estimate of the equatorial velocity
of the star, $v$, to yield $i_\star$.  \citet{2010ApJ...719..602S} used
an empirical relation for Sun-like stars to evaluate $v$ from their
masses and ages.  Alternatively, one can determine the stellar spin
period photometrically from periodic variations in the lightcurve due to
the stellar activity, and then estimate $v$ assuming the stellar radius
\citep[e.g.,][]{Hirano2012}.

The other is asteroseismology \citep{Unno1989, Aerts2010} for which the
key principles are described in detail later, but briefly summarised
below. Thanks to space-borne instruments such as MOST
\citep{2003PASP..115.1023W}, CoRoT \citep{Baglin2006a,Baglin2006b} and
\emph{Kepler} \citep{Borucki2010}, asteroseismology now opens a
good opportunity to unveil the internal structure of many stars
with high precision. This is made possible through the
detection of oscillation modes propagating throughout the stars with
unprecedented precision, because of extraordinary low noise level
and uninterrupted extremely long-term data monitoring with short
sampling cadences, both of which are never available from the
ground-based observations \cite[e.g.][]{Appourchaux2008, Metcalfe2012,
Gizon2013}.  More details about the recent development in
asteroseismology may be found in recent conference proceedings
such as \citet{2012ASPC..462.....S}, \citet{2013ASPC..479.....S}, and
\citet{2014IAUS..301.....G}.  When coupled with non-seismic observables,
the asteroseismic observational information promises accurate
inference of fundamental properties of host stars
\citep[e.g.][]{Bazot2005, Carter2012}.

The stellar rotation affects the frequency spectrum of stellar
oscillation modes.  It induces a multiplet fine structure, whose
frequency separation is dependent on the internal rotation profile of
the star as well as the stellar structure, for each mode.  More
importantly, the apparent profiles of the rotationally induced frequency
multiplets are very sensitive to the inclination angle of the
stellar rotation axis with respect to the line-of-sight, $i_\star$.
In turn, one can infer $i_\star$ quite well from
asteroseismology.

One might wonder how commonly stellar oscillations that enable
asteroseismology can be detected among the host stars of exoplanet
systems. Remember that transiting planet hunting
preferentially select stars with small radii (i.e. low-mass stars
in the main sequence) in order to increase the relative transit
depth in the photometric lightcurve.  Moreover, such low-mass
stars are suitable for the radial velocity follow-up not only because they are
more affected by orbital motion of planets but also because they have
sharp and narrow absorption lines due to their slow spin rotation
velocity. Such low-mass, cool stars have a thick convective envelope
(as in the case of the Sun) that sustains pulsations.  Turbulent motion
with speeds close to that of sound near the stellar surface
stochastically generates acoustic waves, which propagate inside the star
until they are damped. The oscillations with frequencies close to
those of eigenmodes of the star are eventually sustained as many
acoustic modes.  Therefore cool (\emph{i.e.} $\lessapprox 7000$ K)
host stars for exoplanets should commonly exhibit solar-like
oscillations, and thus consitute good targets for asteroseismology.

In this paper, we focus on two specific exoplanetary systems,
HAT-P-7 and Kepler-25; HAT-P-7 is the first example of a system
hosting a retrograde or a polar-orbit planet, while Kepler-25 is a
multi-transiting system with three planets, making them two interesting
examples.  We show that joint analyses of asteroseismology,
transit lightcurve, and the RM effect provide stringent orbital parameter
estimates.

This paper is organised as follows. Section \ref{sec:previous}
summarizes the previous RM measurements and radial velocity (RV) data of
the two systems.  Section \ref{sec:asteroseismology} presents a
detailed description of the basic principle of asteroseismology,
followed by our main results of asteroseismology for the two stars in
Sections \ref{sec:asteroseismology:HATP7} and
\ref{sec:asteroseismology:K25}. Sections \ref{sec:hat-p-7} and
\ref{sec:kepler-25} analyze the {\it Kepler} transit lightcurves and the
RV anomaly of the RM effect, using the asteroseismology results as a
prior, and show how the joint analysis improves the estimate{s} of
the system parameters.  Section \ref{sec:summary} is devoted to further
discussion, and Section \ref{sec:9} summarises the present paper.

\section{Previous Spin--Orbit Measurements
\label{sec:previous}}

\subsection{HAT-P-7
\label{subsec:previous-hatp7}}

The HAT-P-7 system comprises a bright (\emph{V}=10.5) F6 star and a hot
Jupiter transiting the host star with a 2.2-d period \citep[hereafter
P08]{Pal2008}. In addition to the significant spin--orbit misalignment
first revealed by the Subaru spectroscopy \citep{Narita2009, Winn2009},
the fact that the system is in the {\it Kepler} field makes it very
attractive as an asteroseismology target.

Interestingly, there have been three independent measurements of the RM
effect for the HAT-P-7 system, which all indicate the significant {spin--orbit}
misalignment, but do not agree quantitatively.  \citet{Winn2009}
(hereafter W09) performed the joint analysis of the spectroscopic and
photometric transit of HAT-P-7b to obtain $\lambda = \timeform{182D.5}
\pm \timeform{9D.4}$.
For RVs, they analyzed 17 spectra observed with the High Resolution
Spectrograph (HIRES) on the Keck I telescope as well as 69 spectra
observed with the High Dispersion Spectrograph (HDS) on the Subaru
telescope.  Eight of the HIRES spectra were from P08 and taken in 2007,
while the other nine were obtained in 2009.  Among 69 HDS spectra, 40
were obtained on 2009 July 1 that spanned a transit.

On the other hand, \citet{Narita2009} (hereafter N09) determined
$\lambda = \timeform{227D.4} {+\timeform{10D.5} \atop -\timeform{16D.3}}$
(equivalently $\lambda=- \timeform{132D.6} {+\timeform{10D.5}
\atop -\timeform{16D.3}}$) based on the eight HIRES RVs from P08
and 40 HDS spectra spanning the transit on 2008 May 30.  Although they
fixed the transit parameters in the analysis of the RM effect, the
systematics from the uncertainties of these parameters do not seem to
explain the mild discrepancy with the W09 result, according to their
discussion (see cases 1 to 4 in section 4 of N09).

Later on, \citet{2012ApJ...757...18A} (hereafter A12) reported another
measurement of the RM effect, resulting in $\lambda = \timeform{155D}
\pm 14^\circ$.  They analyzed 49 HIRES spectra spanning a transit on the
night 2010 July 23/24 with the priors on transit parameters and
ephemeris from the {\it Kepler} lightcurves.

In this paper, we use the same RV data published in each of the three
papers.  Since the origin of the possible discrepancy in $\lambda$ is
not clear, we analyze each data set separately instead of
combining the three.

\subsection{Kepler-25
\label{subsec:previous-kepler25}}

The Kepler-25 system is one of the few multi-transiting planetary
systems with constrained $\lambda$. It consists of a relatively
bright ($K_{\rm p}=10.7$) host star, two short-period Neptune-sized
planets confirmed with Transit Timing Variations (TTVs)
\citep{2012MNRAS.421.2342S}, and one outer non-transiting planet
detected in long-term RV trend \citep{2014ApJS..210...20M}.
\citet{Albrecht2013} (hereafter A13) measured $\lambda = 7^\circ \pm
8^\circ$ for the larger transiting planet Kepler-25c based on the HIRES
spectra observed for two nights (2011 July 18/19 and 2012 May
31/June 1).  Since the signal-to-noise ratio of the RV anomaly was small
due to the relatively small radius of Kepler-25c, they also analyzed the
time-dependent distortion of the spectral lines directly [known as the
``Doppler shadow'' method; see \citet{2010MNRAS.403..151C}] and obtained
a consistent result, $\lambda= -\timeform{0D.5} \pm \timeform{5D.7}$.

In this paper, we analyze the RVs around the above two transits from A13
alone because our focus is the determination of $\psi$.

\section{Asteroseismology \label{sec:asteroseismology}}

\subsection{Setting Up the Problem }

Due to its sensitivity to the stellar internal structure,
asteroseismology can achieve high-precision determinations of
stellar fundamental parameters \citep{Lebreton2009} (e.g.,
uncertainties of a few percent level for their mass and
radius). The stellar modelling using seismic observables mostly relies
on the stellar pulsation frequencies, usually extracted from the
analysis of the power spectrum of the stellar lightcurve.

For a spherically symmetric star, each eigenmode is characterised
by three quantum numbers; the angular degree $l$, the azimuthal
order $m$ ($-l \le m \le +l$), and the radial order $n$. The degree $l$
corresponds to the number of nodal surface lines, while the
azimuthal order $m$ specifies the surface pattern of the eigenfunction,
with $|m|$ being the number of longitude lines among the $l$ nodal
surface lines. The radial order $n$ corresponds to the number of nodal
surfaces along the radius.  For a non-rotating star, both the radial
eigenfunction and the frequency of each mode are independent of $m$ and
show the $(2l+1)$-fold degeneracy.  The eigenfrequency $\nu$
depends on $l$ and $n$ alone.  Frequencies of high order, acoustic
(or p-) modes of the same low degree $(n \gg l \sim 1)$ are almost
equally spaced and separated on average by a frequency spacing
$\Delta\nu$:
\begin{equation}
\nu(n,l) = \Delta\nu \left(n+{{l}\over{2}}+\alpha \right) 
+ \varepsilon_{n,l},
	\label{eq:2}
\end{equation}
where $\alpha$ is a constant of order unity, and $\varepsilon_{n,l}$ is
a small correction.  The spacing is related to the sound
velocity inside the star by
\begin{equation}
\Delta\nu = \left(2\int_0^{R_\star} {{1}\over{c(r)}}\,dr\right)^{-1}
	\label{eq:3}
\end{equation}
and is sensitive to the mean stellar density $\rho_\star$.
Therefore, knowing the solar density $\rho_\odot=(1.4060 \pm
0.0005)\times 10^3$\,kg\,m$^{-3}$ and its frequency spacing
$\Delta\nu_\odot = 135.20 \pm 0.25$ $\mu$Hz\footnote{From frequencies of
\cite{Garcia2011b}.}, one can estimate the mean stellar density
from the scaling:
\begin{equation}
	\rho_{\star,\mathrm{s}} = \rho_\odot (\Delta\nu/\Delta\nu_\odot)^2. 
	\label{eq:3b}
\end{equation}

The stellar rotation lifts the degeneracy among non-radial modes
($l\neq 0$), revealing a fine structure of modes identified by their
azimuthal order $m$.  In the case of solar-like oscillations, acoustic
modes are excited stochastically by turbulent convection.  This
mechanism is expected to generate almost the same amplitudes in
the rotationally split modes with the same $l$ and $n$.  If this is
the case, in disk-integrated photometry as achieved by \emph{Kepler},
the height of the azimuthal modes in the power spectrum is
sensitive to the stellar inclination angle $i_\star$ due to a
geometrical projection effect [see \cite{Gizon2003} for more details]
and has been widely used to evaluate $i_\star$
\cite[e.g.][]{Benomar2009b, Appourchaux2012}. In turn, it enables us
to measure $\psi$ of exoplanets \citep{Chaplin2013,
2014ApJ...782...14V}, and indeed revealed a significant
spin--orbit misalignment for a red-giant host star system, Kepler-56
\citep{huber2013}.  This dependence of visibility in the power spectrum
is expressed in terms of
\begin{equation}
\label{eq:legendre}
\mathcal{E}(l,m, i_\star)=\frac{(l-|m|)!}{(l+|m|)!} \left[P^{|m|}_l (\cos i_\star)\right]^2 ,
\label{eq:4}
\end{equation}
where $P^{|m|}_l$ is the associated Legendre function and the integral
of $\mathcal{E}(l,m, i_\star)$ over $\cos i_\star$ is normalised by
$(2l+1)^{-1}$.

Solar-like oscillators such as Kepler-25 and HAT-P-7 are typically slow
rotators for which the centrifugal force can be neglected. In addition,
there is no evidence of a strong magnetic field and we can safely neglect
it \citep{Reese2006, ballot2010}.  If the internal
rotation of the star is independent of the latitude and the longitude,
the split frequencies are simply written as
\begin{equation}
\nu(n,l,m) = \nu(n,l)  + m \,\delta\nu_{\rm s}(n,l),
	\label{eq:5}
\end{equation}
where $\delta\nu_{\rm s}(n,l)$ is the rotational splitting [e.g. \cite{Appourchaux2008, Benomar2009, Chaplin2013}].

\begin{figure*}[t]
  \begin{center}
	\subfigure{\epsfig{figure=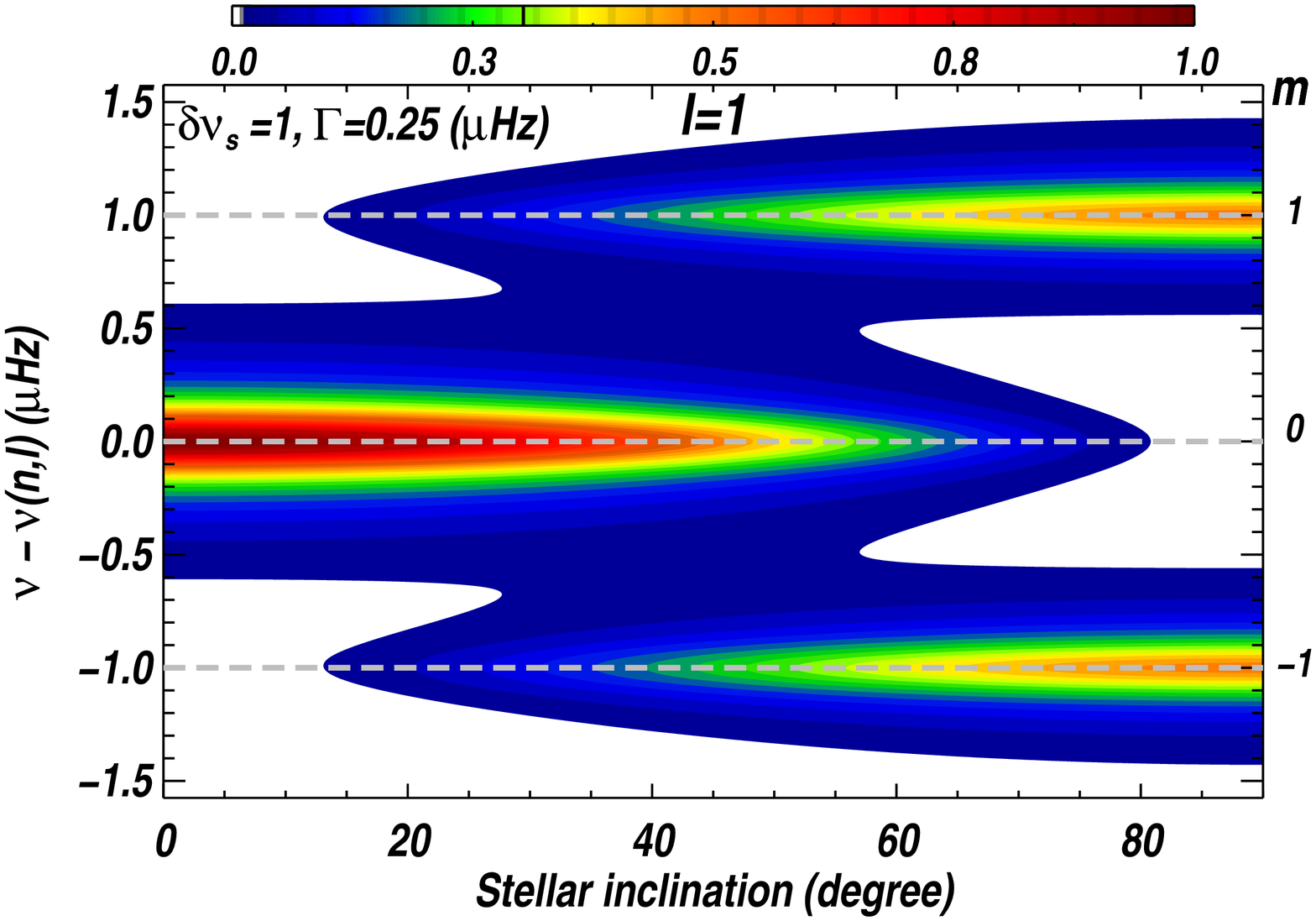, angle=90, width=8cm, height=6cm}}
	\subfigure{\epsfig{figure=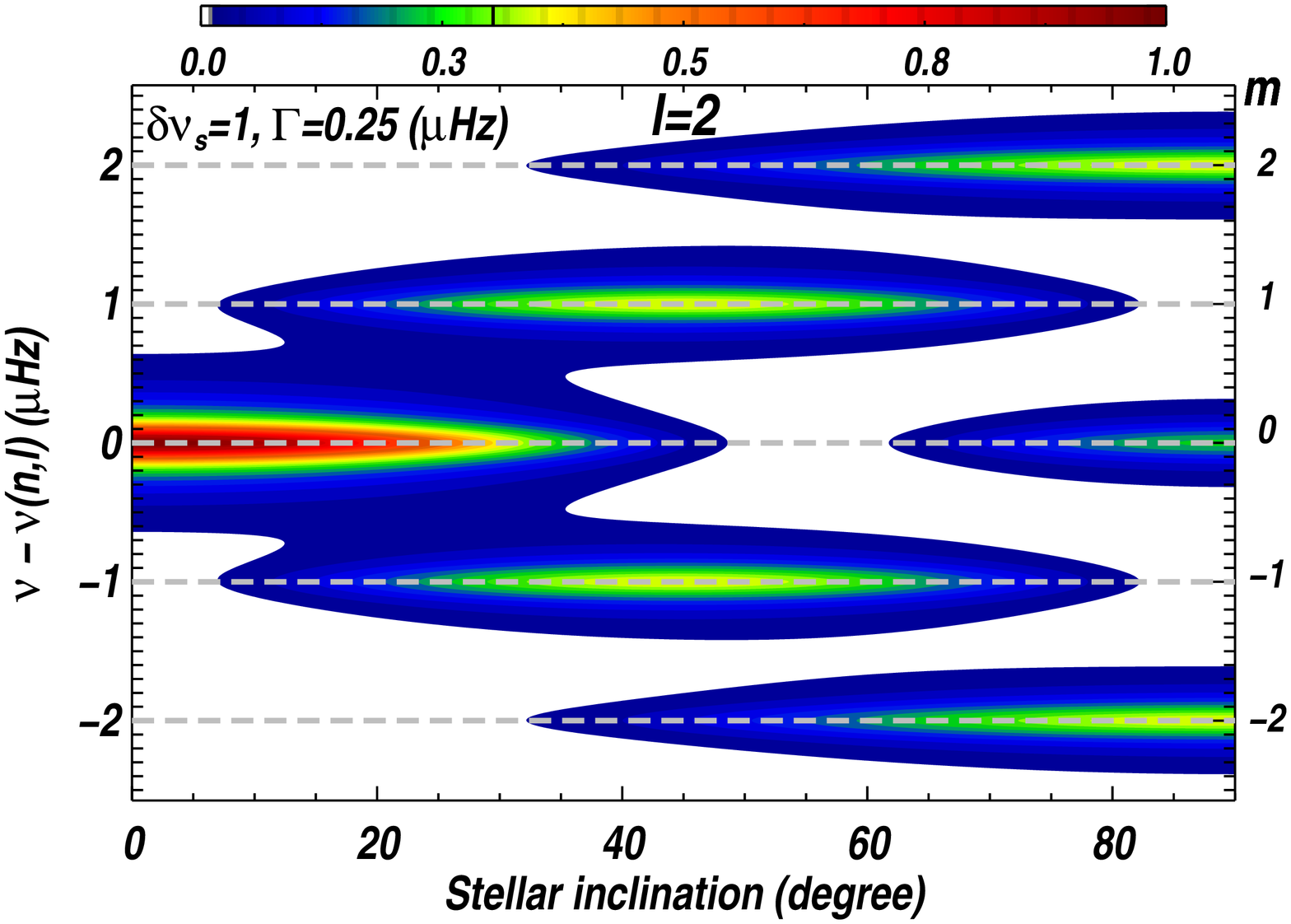, angle=90, width=8cm, height=6cm}}
  \end{center}
\caption{The figures illustrate the dependence between stellar inclination and rotation. The relative power of azimuthal components $m$ for $l=1$ and $l=2$ in the ideal case 
($\delta\nu_s \gg \Gamma$) is indicated by the color scale and is calculated using Equation \ref{eq:power}.
Horizontal dash lines indicate frequencies of the multiplets $m$. Each inclination is characterised by a unique $l=1$ and $l=2$ mode structure. This property is used to infer the stellar inclination. Zero degree corresponds to a star seen from a pole, 90 degree to a star seen from the equator.}  
\label{fig:is-nus-ideal}
\end{figure*} 
\begin{table*}
\caption{Non-seismic observables of HAT-P-7 and Kepler-25. All but $v\sin i_\star$ are used for stellar modelling.}
\begin{center}
\begin{tabular}{cccrcc|l}
\toprule
Star &  $L/L_\odot$ & $\log g$ (cgs) & [Fe/H] &  $T_{\rm eff}$ (K) & $v\sin i_\star$ (km\,s$^{-1}$) & Source \\
\midrule
HAT-P-7 & $4.9 \pm 1.1$ & $4.070 \pm 0.06$ & $0.26 \pm 0.08$ & $6350 \pm 80$ & $3.8 \pm 0.5$ & \cite{Pal2008}  \\
Kepler-25 & N/A & $4.278 \pm 0.03$ & $-0.04 \pm 0.10$ & $6270 \pm 79$ & $9.5 \pm 0.5$ & \cite{2014ApJS..210...20M}\\
\bottomrule
\end{tabular} 
\end{center}
\label{tab:nonseism_obs} 
\end{table*}
\renewcommand{\arraystretch}{0.7} 
\begin{table}
\caption{Detected pulsation frequencies of HAT-P-7 and Kepler-25. The radial order $n$ is determined by the best models matching observables.}
\begin{center}
\begin{tabular}{c|crc|crc}
\toprule
          & \multicolumn{3}{c}{HAT-P-7} & \multicolumn{3}{c}{Kepler-25} \\ 
$l$ & $n$ & $\nu_{n, l, m=0}$ &  $\sigma$ & $n$ & $\nu_{n, l, m=0}$ &  $\sigma$  \\
        &     &    ($\mu$Hz) &  ($\mu$Hz) &  & ($\mu$Hz) &  ($\mu$Hz) \\
\midrule
0  &  11 & 715.50   & 0.30  & 16 & 1691.79   & 0.52 \\
0  &  12 & 771.57   & 0.51  & 17 & 1788.58   & 0.23 \\
0  &  13 & 828.34   & 0.30  & 18 & 1884.39   & 0.36 \\
0  &  14 & 885.91   & 0.26  & 19 & 1981.33   & 0.18 \\
0  &  15 & 944.88   & 0.25  &  20 & 2080.08   & 0.32 \\
0  &  16 & 1004.77  & 0.22  & 21 & 2178.68   &  0.43 \\
0  &  17 & 1064.83  & 0.20  & 22 & 2277.00   &  0.32  \\
0  &  18 & 1123.17  & 0.23  & 23 & 2375.48   & 0.67  \\
0  &  19 & 1181.90  & 0.23  & 24 & 2472.91   & 0.59  \\
0  &  20 & 1240.53  & 0.27  & 25 & 2570.03   & 1.43  \\
0  &  21 & 1300.53  & 0.35  &                 &         \\   
0  &  22 & 1360.78  & 0.43  &                 &         \\  
0  &  23 & 1421.55  & 0.94   &                 &         \\
0  &  24 & 1482.03  & 0.75   &                 &         \\
0  &  25 & 1542.96  & 1.21   &                 &         \\
\midrule
1  &  11 & 740.79    & 0.22	  &  16 & 1736.27  &     0.79 \\
1  &  12 & 796.71    & 0.35   & 17 & 1832.49   &    0.20  \\
1  &  13 & 854.00    & 0.23   & 18 & 1929.17   &   0.28  \\
1  &  14 & 911.89    &  0.20  &  19 & 2026.97   &   0.28  \\
1  &  15 & 971.85    & 0.16   &  20 & 2125.46   &    0.32  \\
1  &  16 & 1031.54  &  0.15  &  21 & 2224.32   &    0.51  \\
1  &  17 & 1091.15  &  0.15  &  22 & 2323.04   &   0.32  \\
1  &  18 & 1149.92   & 0.17  &   23 & 2421.68   &   0.53  \\
1  &  19 & 1208.36   &  0.17 &   24 & 2521.29   &   0.63  \\
1  &  20 & 1267.82   &  0.23 &   25 & 2621.12   &   1.12  \\
1  &  21 & 1327.41   &  0.27 &                   &           \\  
1  &  22 & 1388.49   &  0.36 &                   &           \\
1  &  23 & 1448.96   &  0.46 &                   &           \\
1  &  24 & 1509.40   &  0.54 &                   &           \\
1  &  25 & 1569.30   &  0.92 &                   &           \\ 
\midrule
2  &  10 & 710.81     &  0.63 &  15 & 1683.26    &    3.88 \\
2  &  11 & 767.31     &  0.62 &  16 & 1779.57   &      2.17 \\
2  &  12 & 824.46     &  0.53 &  17 &  1875.12   &     1.39  \\
2  &  13 & 882.27     &  0.54 &  18 &  1972.55   &     0.67  \\
2  &  14 & 940.46     &  0.34 &  19 &  2071.55   &     0.77  \\
2  &  15 & 1000.17   &  0.49 &  20 &  2170.64   &     0.88  \\
2  &  16 & 1059.82   &  0.35 &  21 & 2269.90   &     1.14  \\
2  &  17 & 1118.74    &  0.31 &  22 & 2368.62   &     1.16  \\
2  &  18 & 1177.89    &  0.39 &  23 &  2467.03   &     1.38  \\
2  &  19 & 1236.33    &  0.42 &  24 &  2565.48   &     2.79  \\
2  &  20 & 1296.40    &  0.50 &                   &           \\ 
2  &  21 & 1356.39    &   0.50 &                   &           \\
2  &  22 & 1417.09    &   0.97  &                   &           \\
2  &  23 & 1478.41    &   0.97 &                   &           \\
2  &  24 & 1539.79    &   1.50 &                   &           \\ 
\bottomrule
\end{tabular} 
\end{center}
\label{tab:seism:freq} 
\end{table}
\renewcommand{\arraystretch}{1.0} 

It should be noted that because of the modes stochastic nature, each solar-like mode has a Lorentzian profile
in the power spectrum \citep{harvey1985}. Thus, the stellar oscillations can be expressed as a sum of Lorentzian over $n$, $l$ and $m$,
\begin{equation}
\label{eq:power}
P(\nu)= \sum_{n,l}\sum_{m=-l}^l 
\frac{\mathcal{E}(l,m, i_\star)H(n,l)}
{1+4(\nu-\nu(n,l,m))^2/\Gamma^2(n,l,m)} .
\end{equation}
Each mode is therefore not only characterised by its frequency $\nu (l,n,m)$, but also by a height $H(n,l,m)=\mathcal{E}(l,m, i_\star)
H(n,l)$ and a full width at half maximum $\Gamma(n,l,m)$ (hereafter
called width). Here, $H(n,l)$ is the intrinsic height for the mode of
$n$ and $l$.  The heights and the widths of the modes retain information
on, for example, the modes excitation mechanism and on non-adiabatic
processes.

In order to show the sensitivity of the asteroseismology analysis
to $i_\star$, we plot $P(\nu)$ in Figure \ref{fig:is-nus-ideal} as a
function of $i_\star$ and $\nu-\nu(n,l)$.  The left and right panels
correspond to $l=1$ and $l=2$ modes, respectively. The plot is
color-coded according to the amplitude of $P(\nu)$ and for a given $n$ ({\it i.e,} a sum of
the $(2l+1)$ Lorentzian profiles).  Figure \ref{fig:is-nus-ideal} is
presented simply for illustrative purpose, and is computed from
equations (\ref{eq:legendre}) and (\ref{eq:power}), assuming
$\delta\nu_{\rm s} = 1\,\mu$Hz and $\Gamma=0.25\mu$Hz.  The condition
$\delta\nu_{\rm s} \gg \Gamma$ breaks the degeneracy among the
rotationally split $m$ components.  As demonstrated in this figure, in
the case of $i_\star \simeq 0^\circ$, that is, when we see the star from
the pole, only the $m=0$ component is visible as a singlet for both of
$l=1$ and $l=2$ modes. On the other hand, in the case of $i_\star \simeq
90^\circ$, the rotational splitting appears as a doublet in the case of
$l=1$ and as a triplet in the case of $l=2$.  Thus for a given value of
$i_\star$, the power is the result of a unique configuration of height for the
$m$ components, which enables us to infer the value of $i_\star$ from the
$l=1$ and $l=2$ mode profiles. Note, however, that because Equation
(\ref{eq:legendre}) depends on $|m|$ 
, solutions in the four quadrants of the trigonometric circle are degenerate and one cannot distinguish
between $i_\star$ and $(180^{\circ}-i_\star)$.  
 
\subsection{Data processing and modeling 
\label{sec:asteroseismology:method}}

The {\it Kepler} Space Telescope collected time series lightcurves of
about 160,000 stars over the 115 square degrees field-of-view from
its 372.5-d, heliocentric Earth-trailing orbit over its four-year
lifetime for 2009 -- 2013.  Its major purpose was to find
extra-solar planets by detecting a small amount drop of the visual
brightness of their parent stars, caused by the transits of the planets
in front of the stars.  So the photometric asteroseismology and planet
studies are synergistic.  Four times per orbit the satellite was
scheduled to perform a roll to keep its solar panels facing the Sun, so
the data were divided into `Quarters' (1/4 of its 372.5-d heliocentric
orbit), denoted as Q$n$.

For HAT-P-7, we use Q0 to Q16 (1437 days in total) of
\emph{Kepler} data taken every 1-min (`Short Cadence' data; SC), while
for Kepler-25, we used Q5 to Q16 (1114 days) SC data.  After
removing the transits from the lightcurve with a median
high-pass filter of an adequate frequency width, we compute the
power spectrum of each star following the method described in
\cite{Garcia2011}. The high-pass filter is efficient to remove the
signal of the transit in the power spectrum without altering the stellar
pulsation characteristics, since the orbital periods of the detected
planets around HAT-P-7 and Kepler-25 are of the order of days, while
stellar pulsation periods are in the minute range. To extract the
mode parameters, we perform a Lorentzian profile fit to each mode
that exhibits significant power. We use a Markov Chain Monte Carlo
(MCMC) method and a similar method to \cite{Benomar2009} but with a
smoothness condition on the frequencies. \citep[see][]{Benomar2013a, Benomar2014}.

The prior on the rotational splitting $\delta\nu_{\rm s}$ is uniform
between 0 and 8\,$\mu$Hz. The prior on $i_\star$ is chosen to be
uniform in $\cos i_\star$ for $0<\cos i_\star<1$, and is equivalent to
the random uniform distribution of $i_\star$.  Because of the symmetries
in Equation (\ref{eq:legendre}), we only consider solutions of $0^\circ
\leq i_\star \leq 90^\circ$ in what follows.

\begin{figure*}
  \begin{center}
	\includegraphics[width=0.55\linewidth,angle=90]{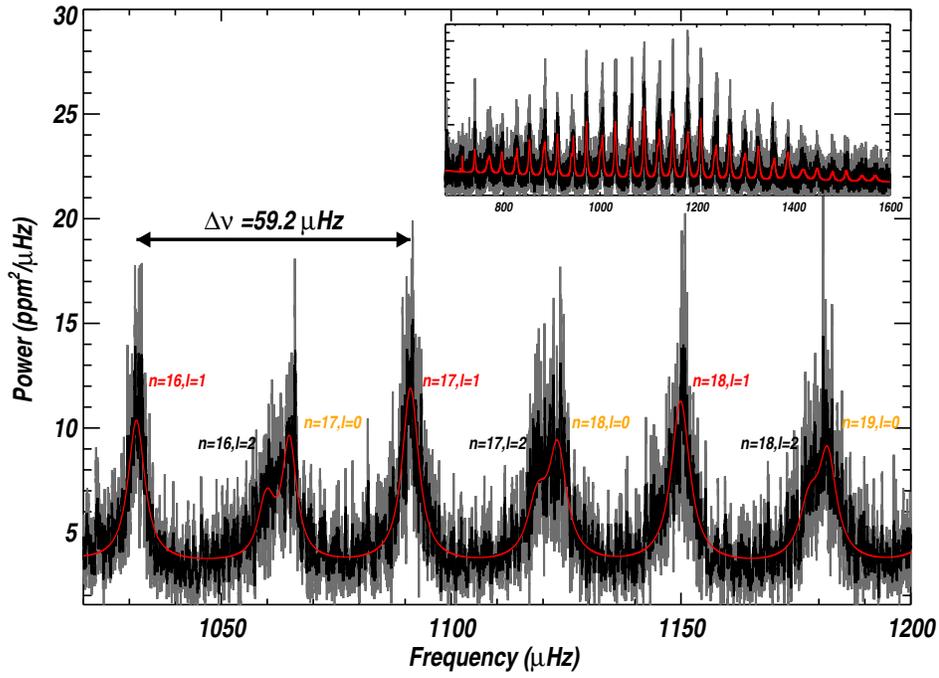}
  \end{center}
\caption{\textbf{HAT-P-7.} Power spectrum over three radial orders for
modes with highest signal-to-noise ratio. The spectrum is shown after a
boxcar smooth over $0.08$ $\mu$Hz (grey) and $0.24$ $\mu$Hz (black). The
best fit is the solid red line. The inset shows all the extracted
modes.}  \label{fig:spectrum:Hatp7}
\end{figure*}

\begin{figure*}
  \begin{center}
	\includegraphics[width=0.55\linewidth,angle=90]{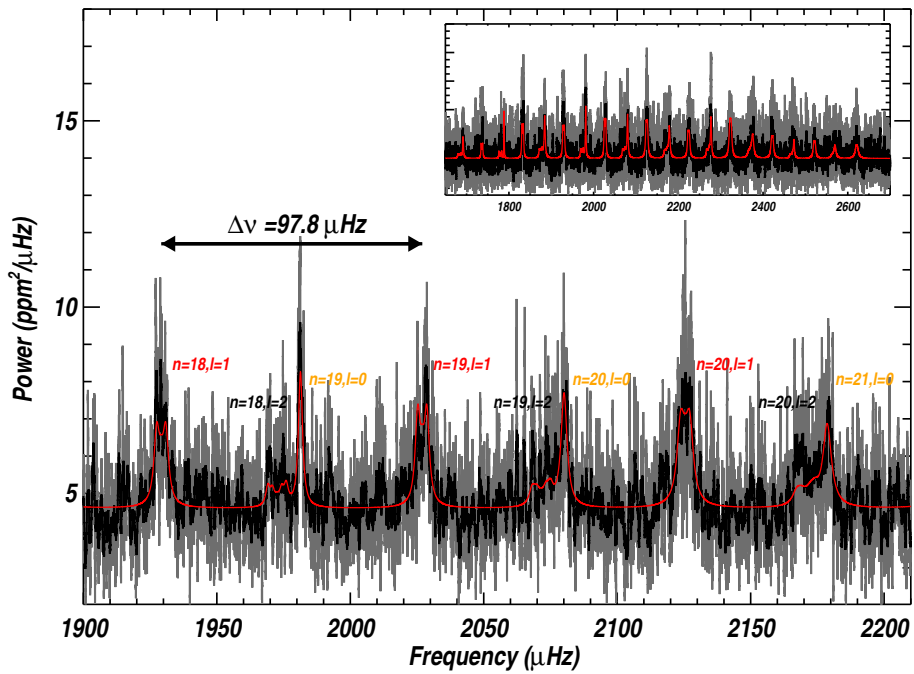}
  \end{center}
\caption{\textbf{Kepler-25}. Power spectrum over three radial orders for
modes with highest signal-to-noise ratio. The spectrum is shown after a
boxcar smooth over $0.21$ $\mu$Hz (grey) and $0.83$ $\mu$Hz (black). The
best fit is the solid red line. The inset shows all the extracted
modes.}  \label{fig:spectrum:K25}
\end{figure*}

Figures \ref{fig:spectrum:Hatp7} and \ref{fig:spectrum:K25} show the
resulting power spectra and their best-fit models for HAT-P-7 and
Kepler-25, respectively.  The identified pulsation modes and the derived
pulsation frequencies of the central component of multiplets are listed
in Table\,\ref{tab:seism:freq}.


Stellar models that simultaneously match non-seismic observables
(cf. Table \ref{tab:nonseism_obs}) and seismic observables (frequencies
in Table \ref{tab:seism:freq}) are found using the `astero' module of
the Modules for Experiments in Stellar Astrophysics (MESA) evolutionary
code \citep{paxton2011, paxton2013}.  Stellar models are calculated
assuming a fixed mixing length parameter $\alpha_{\rm MLT}=2.0$
and an initial hydrogen abundance $X=0.7$. The opacities are
calculated using the MESA standard equation of state from the
opacity table in \cite{Asplund2009}. These are applicable for
stars with effective temperature $10^{3}\,\rm K< T_{\rm eff} <
10^{4.1}\,\rm K$.  Nuclear reactions are set to include standard hydrogen and
helium burning; the pp-chain and the CNO cycle in addition to the triple
alpha reaction. 

It is known that semiconvective zones are present in stars of $\sim
1.5\,M_\odot$. Since a small convective core in such stars expands due
to the growing importance of the CNO cycle, the opacity is larger at the
outer side of the convective boundary than at its inner side.  We adopt
the M. Schwarzschild treatment to define the boundary between the
convective and radiative zones in such a case. With expected mass larger
than 1.2\,$M_\odot$, HAT-P-7 and Kepler-25 may have a convective
core. Then, the nature of the transition (e.g. sharp or smooth)
between convective and radiative regions may have a significant impact
on the seismic frequencies \citep[e.g.][]{Monteiro1994}. Thus, to
describe a possible extension of the convective zone inside the
radiative zone, we have included an overshoot. Diffusion was not
implemented.  Mass $M_\star$, metallicity [Fe/H], helium abundance $Y$,
the coefficient for overshooting $\alpha_{\rm ov}$, and age are
treated as free parameters.

Eigenfrequencies are calculated assuming adiabaticity and using {\tt
ADIPLS} \citep{JCD2008b}. We apply surface effect corrections to the
frequencies, following the method of \cite{Kjeldsen2008}.  The search
for the best model involves a simplex minimisation approach
\citep{Simplex} using the $\chi^2$ criteria. Uncertainties are then
estimated by evaluating the $\chi^2$ for solutions surrounding the best
model and by weighing the model parameters with Likelihood $\propto
\exp(-\chi^2/2)$.

\subsection{Mode Degree Identification 
\label{sec:asteroseismology:modeI}}

Prior to modelling a star, it is important to identify the degree $l$ from the power spectrum.
In solar-like cool stars (K, G type) the
identification is often obvious and relies on the \'echelle diagram
\citep{Grec1983}. An \'echelle diagram is built by dividing the power spectrum into frequency bins of interval $\Delta$, that are stacked in order to form an image in which the power is color-coded. In this image, the Y-axis represents the central frequency of each bin, while the X-axis corresponds to the frequency modulo $\Delta$. Note that the central frequency of the bins is a discrete quantity and one could use instead an integer for the Y-axis.
Figures \ref{fig:EDs-HATP7} and \ref{fig:EDs-K25} are the corresponding
\'echelle diagram for HAT-P-7 and Kepler-25 stacked with
$\Delta=59.9\,\mu$Hz and $\Delta=97.8\,\mu$Hz, respectively.  If we
choose $\Delta=\Delta\nu$, the excess power due to the modes of the same
degree $l$ should show up along an almost straight vertical
line\footnote{For HAT-P-7, $\Delta$ is chosen slightly different than
$\Delta\nu$ for a better rendering of the \'echelle diagram.}.  This is
because p-modes of the same degree are almost regularly spaced in
frequency, as implied by Equation (\ref{eq:2}).

Equation (\ref{eq:2}) shows that $\nu({n,l})=\nu({n-1, l+2})$ as long as
$\varepsilon_{n,l}$ is small. Thus the eigenmodes of {($n$, $l=0$)} and
{($n-1$, $l=2$)} have approximately the same frequencies. The same is true
for ($n$, $l=1$) and ($n-1$, $l=3$). On the other hand, the pulsation
amplitude of the surface, and {consequently} the integrated
luminosity variation, {are} smaller for larger $l$ modes.  Thus, the
detected photometric amplitudes of the pulsation are usually dominated
by $l=0$ and $l=1$, and $l \ge 3$ are often buried in the noise. 

This is why the careful visual inspection of the relative height and
frequency of the power spectra {enable us to} identify the corresponding modes.
This approach works for Kepler-25, but not for HAT-P-7 in reality.  The
power spectrum of HAT-P-7 exhibits significant mixture of $l=0$ and
$l=2$ modes, and it is hard to disentangle them by visual inspection.
In such a case that the modes of the same $l$ are almost regularly
spaced in frequency, there exist two possibilities: either (S1) the fit
misidentifies the modes, or (S2) the fit correctly identifies the mode.
As for the former, all modes of degrees $l=0$ and $l=2$ would be
misidentified as $l=1$ modes (and vice-versa).  This problem of mode
identification is recurrent in F stars and was first encountered in a
star observed by CoRoT, HD\,49933 \citep{Appourchaux2008}.

The most likely solution among the two competitive solutions (S1) and
(S2) described above may be judged by the Bayes factor between S1
and S2 [see \cite{Benomar2009, Benomar2009b, Appourchaux2012} for more
details]. Using our MCMC samples, we evaluated the Bayes factor at
$10^6:1$ in favour of modes with frequencies listed in Table
\ref{tab:seism:freq}.  According to \cite{Jeffreys1961} , the Bayes
factor $>100$ is ``Decisive'', and thus one can safely assume that the
mode identification is correct. We also note that use of the empirical
approach detailed in \cite{White2012} reproduces the same degree
identification.

Furthermore, there is not clear evidence for $l=3$ in the \'echelle
diagram. To verify this quantitatively, we attempted to detect modes of
degree $l = 3$ by comparing the Bayes factor between a model $M_{l \leq
3}$, that includes those modes, with a model $M_{l \leq 2}$ that does
not. We obtained a factor $\simeq 2:1$ and $\simeq 2.5:1$ for HAT-P-7
and Kepler-25 respectively, in favour of $M_{l \leq 2}$, which is the
simplest model. Thus modes of degree $l =3$ are not conclusively
detected.

\begin{table}[!h]
\caption{Stellar model characteristics for HAT-P-7 and Kepler-25 derived
with MESA.  $\rho_{\star,\rm m}$ is the density derived from
modelling. $\rho_{\star,\rm s}$ is the density derived by rescaling the
Sun density using the average frequency separation $\Delta \nu$. }
\begin{center}
\begin{tabular}{c|c|c}
\toprule
parameter     & HAT-P-7  & Kepler-25  \\
\midrule
$M_\star$ $(M_\odot)$ & $1.59 \pm 0.03$ &  $1.26 \pm 0.03$  \\
$R_\star$ $(R_\odot)$ & $2.02 \pm 0.01$ & $1.34 \pm 0.01$  \\ 
$[{\rm Fe/H}]$ &  $0.32 \pm 0.04$ & $0.11 \pm 0.03$  \\
$T_{\rm eff}$ (K) & $6310 \pm 15$       &    $6354 \pm 27$  \\
Age (Myrs) &  $1770 \pm 100$ & $2750 \pm 300$  \\
$\alpha_{\rm ov}$ & $0.000 {{+0.002}\atop{-0.000}}$ & $0.007 \pm 0.003$ \\
$L/L_\odot$ &  $5.84 \pm 0.05$ & $2.64 \pm 0.07$  \\
$\log g$ (cgs) & $4.029 \pm 0.002$ & $4.285 \pm 0.003$ \\
$\rho_{\star,\rm m}$ (10$^3$\,kg\,m$^{-3}$) & $0.2708 \pm 0.0035$ &  $0.7367 \pm 0.0137$ \\
$\rho_{\star,\rm s}$ (10$^3$\,kg\,m$^{-3}$)  & $0.2696 \pm 0.0011$ &   $0.7356 \pm 0.0030$ \\
\midrule
reduced $\chi^2$ & $1.73$ & $1.03$ \\
\bottomrule
\end{tabular} 
\end{center}
\label{tab:output_model} 
\end{table}

\section{Asteroseismology of HAT-P-7 
\label{sec:asteroseismology:HATP7}}

The power spectrum of HAT-P-7 (Figure\,\ref{fig:spectrum:Hatp7}) shows a
broad range of modes, spanning over 15 different radial orders
with a high signal-to-noise ratio (Table \ref{tab:seism:freq}),
enabling us to infer modes properties with an unprecedented precision
for an F-star. Our asteroseismic analysis detected a total of 45 modes
of degree $l=0$, $1$ and $2$, for which the frequencies are listed in
Table \ref{tab:seism:freq}.

\subsection{Fundamental Properties}
\begin{figure}[t]
  \begin{center}
	\subfigure{\epsfig{figure=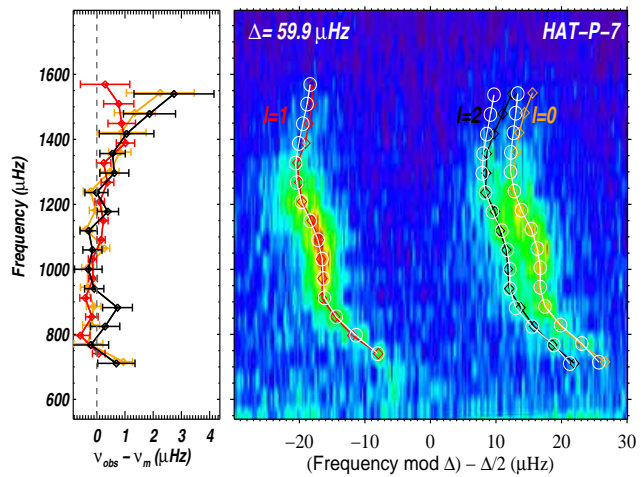, angle=90, width=9.5cm, height=7.5cm}}
  \end{center}
\caption{\textbf{Left panel.} Difference between observed frequencies $\nu_{\rm obs}$ of HAT-P-7
and best model frequencies $\nu_{\rm m}$. $l=0, 1, 2$ are shown as orange, red and
black diamonds respectively. \textbf{Right panel.} \'Echelle diagram
showing the observed power spectrum (background), the observed frequencies (diamonds) and the frequencies from the best model (white circles).}
\label{fig:EDs-HATP7}
\end{figure}

The \'echelle diagram of HAT-P-7 (Figure\,\ref{fig:EDs-HATP7}) shows
clear departures from a straight line, which is mostly the signature of
the transition between the outer convective zone and the radiative
zone. This is because discontinuities within the structure translate
into steep gradients in the acoustic structure of a star, which induce
frequency modulations of periods related to the acoustic depth of the
discontinuities \citep[e.g.][]{Vorontsov1988, Monteiro1994,
Roxburgh2003}. In modelling HAT-P-7, it is therefore important to
find models that match not only the average frequency separation
$\Delta\nu$ (which is sensitive only to the mean density) but also all
individual frequencies accurately.

\begin{figure}[t]
  \begin{center}
	\subfigure{\epsfig{figure=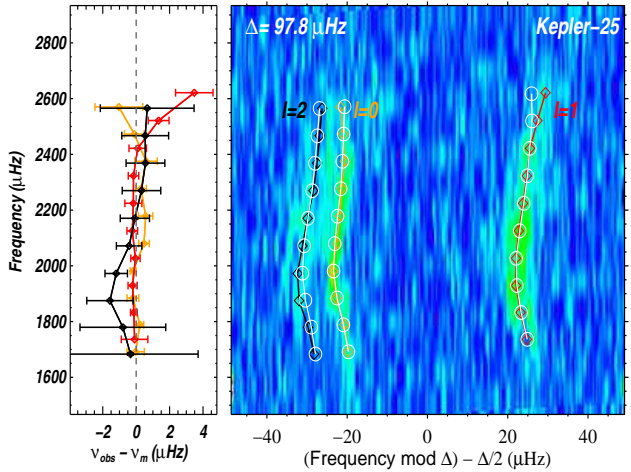, angle=90, width=9.5cm, height=7.5cm}}
  \end{center}
\caption{\textbf{Left panel.} Difference between observed frequencies $\nu_{\rm obs}$ of Kepler-25
and best model frequencies $\nu_{\rm m}$. $l=0, 1, 2$ are shown as orange, red and
black diamonds respectively. \textbf{Right panel.} \'Echelle diagram
showing the observed power spectrum (background), the observed frequencies (diamonds) and the frequencies from the best model (white circles).} 
\label{fig:EDs-K25}
\end{figure} 


 Following the method described in Section
\ref{sec:asteroseismology:method}, we found that the best model implies
$M_\star=1.59 \pm 0.03\,M_\odot$ (cf. Table \ref{tab:output_model} for
main characteristics of the model), which is slightly greater than what
was reported in earlier seismic studies; \cite{JCD2010} used Q0 and Q1
{\it Kepler} data ($\approx 60$ days long) and reported $M_\star = 1.520
\pm 0.036\,M_\odot$. They fitted individual frequencies corrected from
the surface effects \citep{Kjeldsen2008} and they used the {\tt ASTEC}
evolutionary code with a method and physics similar to what we
adopted in the present paper\footnote{Opacity tables and some nuclear
reaction rates are different.}.  \cite{Oshagh2013} carried out an
analysis of HAT-P-7 using {\it Kepler} Q0 to Q2 (144 days long). Their
approach slightly differs from ours as they did a non-adiabatic
frequency calculation. They reported $M_\star=1.415 \pm
0.020\,M_\odot$. Furthermore \cite{VanEylen2012} used {\it Kepler} data
from Q0 to Q11 and reported $M_\star=1.361 \pm 0.021\,M_\odot$. While
our model values are consistent with those quoted in \cite{JCD2010}
within $2\sigma$, the other estimates are significantly different. Thus
we discuss the issue below.

First of all, \cite{JCD2010} and our study result in consistent
mean stellar densities\footnote{Using an MCMC analysis, they found
$M_\star=1.520 \pm 0.036\,M_\odot$ and $R_\star =1.991 \pm
0.018\,R_\odot$, corresponding to $\rho_{\star,\rm m} = (0.2707 \pm
0.0010)\times 10^3$\,kg\,m$^{-3}$, while our model implies
$\rho_{\star,\rm m}= (0.2708 \pm 0.0035)\times 10^3$\,kg\,m$^{-3}$. } at
$1\sigma$.  In contrast, \cite{Oshagh2013} obtain $\rho_{\star,\rm
m}=(0.2778 \pm 0.0059)\times 10^3$\,kg\,m$^{-3}$ and \cite{VanEylen2012}
$\rho_{\star,\rm m}= (0.2781 \pm 0.0017)\times 10^3$\,kg\,m$^{-3}$,
which are consistent within $1\sigma$.  While the differences
between \cite{Oshagh2013} and the present study may be due to the
non-adiabatic treatment of model frequencies and to the data quality as
well, this cannot explain the low mass found by \cite{VanEylen2012}.
Nevertheless, although the model in figure 2 of \cite{VanEylen2012}
has a small value of $\chi^2$, it does not seem to reproduce accurately
their individual frequencies. Moreover their method of measuring the
frequencies differs from ours (frequencies are measured by taking the
frequency at maximum height of a smooth spectrum) and they reported
larger uncertainties than what we obtain here.

In order to see if the difference in methodology could explain the
apparent discrepancies, we looked for the best model (minimum $\chi^2$)
assuming $M_\star=1.36\,M_{\odot}$, to be coherent with
\cite{VanEylen2012}.  The best model has a $\chi^2=24.6$, approximately
14 times higher than the best model shown in Table
\ref{tab:output_model} and does not reproduce accurately the individual
oscillation frequencies.  The mean stellar density $\rho_{\star,\rm
m}=(0.2562 \pm 0.0002)\times 10^3$\,kg\,m$^{-3}$ is also significantly
different. Thus we conclude that mass of $\approx 1.36\,M_{\odot}$
is less favored than $\approx 1.59\,M_{\odot}$, from our seismic
observables.

The best-fit model of the present study implies that the HAT-P-7
has a convective core that extends up to 6.9\% of the stellar radius,
while the outer convective zone represents approximately 13.1\% of the
stellar radius. The central hydrogen abundance $X_{\rm c}=0.214$, which
corresponds to 32\% of its initial core hydrogen, indicates that the
star is at a late stage in its main sequence.  Finally we note that the
best model of HAT-P-7 has no need of surface effect correction.
 
\subsection{Rotation and Inclination}

\begin{figure*}[t]
  \begin{center}
        \subfigure{\epsfig{figure=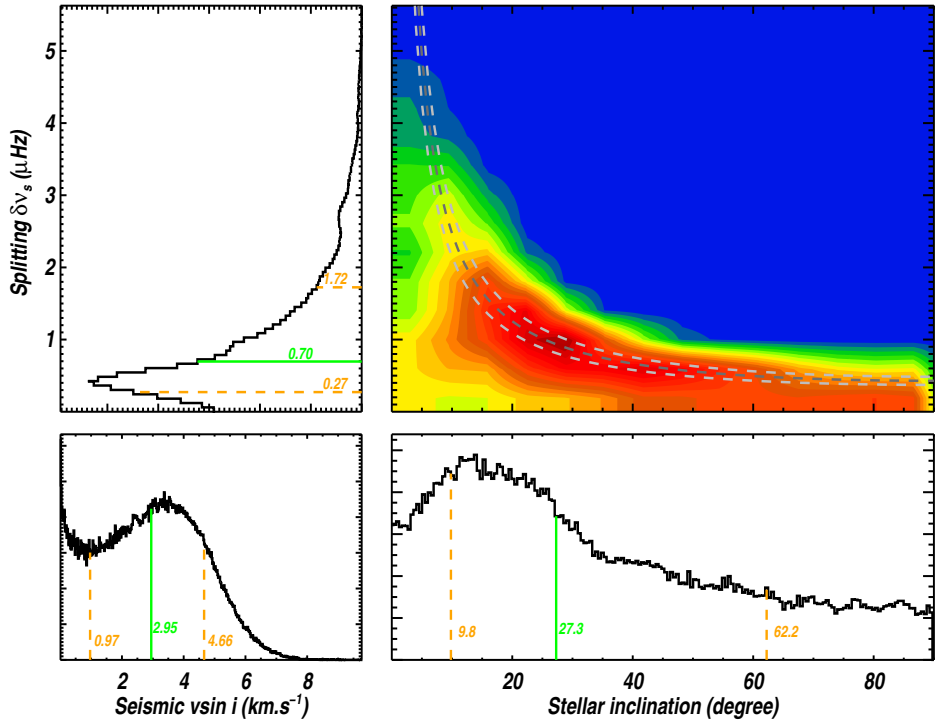, angle=90, width=16cm, height=12cm}}
  \end{center}
\caption{\textbf{Upper right.} Joint posterior probability distribution
of the stellar inclination and the rotation for HAT-P-7. Blue represents
region of lowest probability. Red areas are of highest
probability. Superimposed and using a dark grey dotted line, we show the
spectroscopic $v\sin i_\star$ from P08 with its $1\sigma$ uncertainty
intervals (grey dotted lines). \textbf{Upper left.} Marginalized
probability density function for the rotational splitting. \textbf{Lower
right.} Marginalized probability density function for the stellar
inclination. \textbf{Lower left.} Seismic $v\sin i_\star$, inferred
using the probability density for the rotational splitting, the
inclination and the radius of HAT-P-7.}  \label{fig:is-pdfs-HATP7}
\end{figure*}

\begin{figure*}[t]
  \begin{center}
	\subfigure{\epsfig{figure=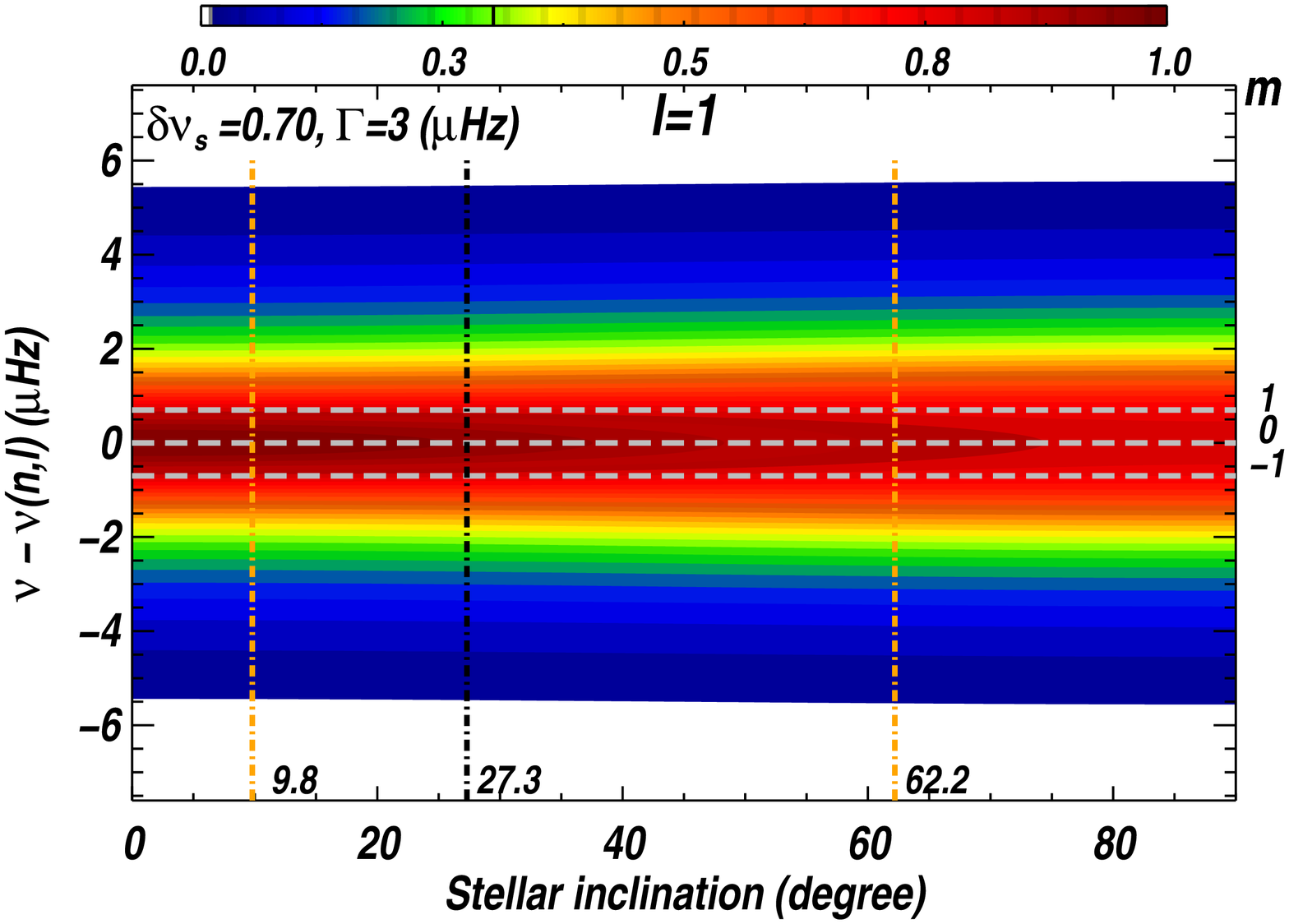, angle=90, width=8cm, height=6cm}}
	\subfigure{\epsfig{figure=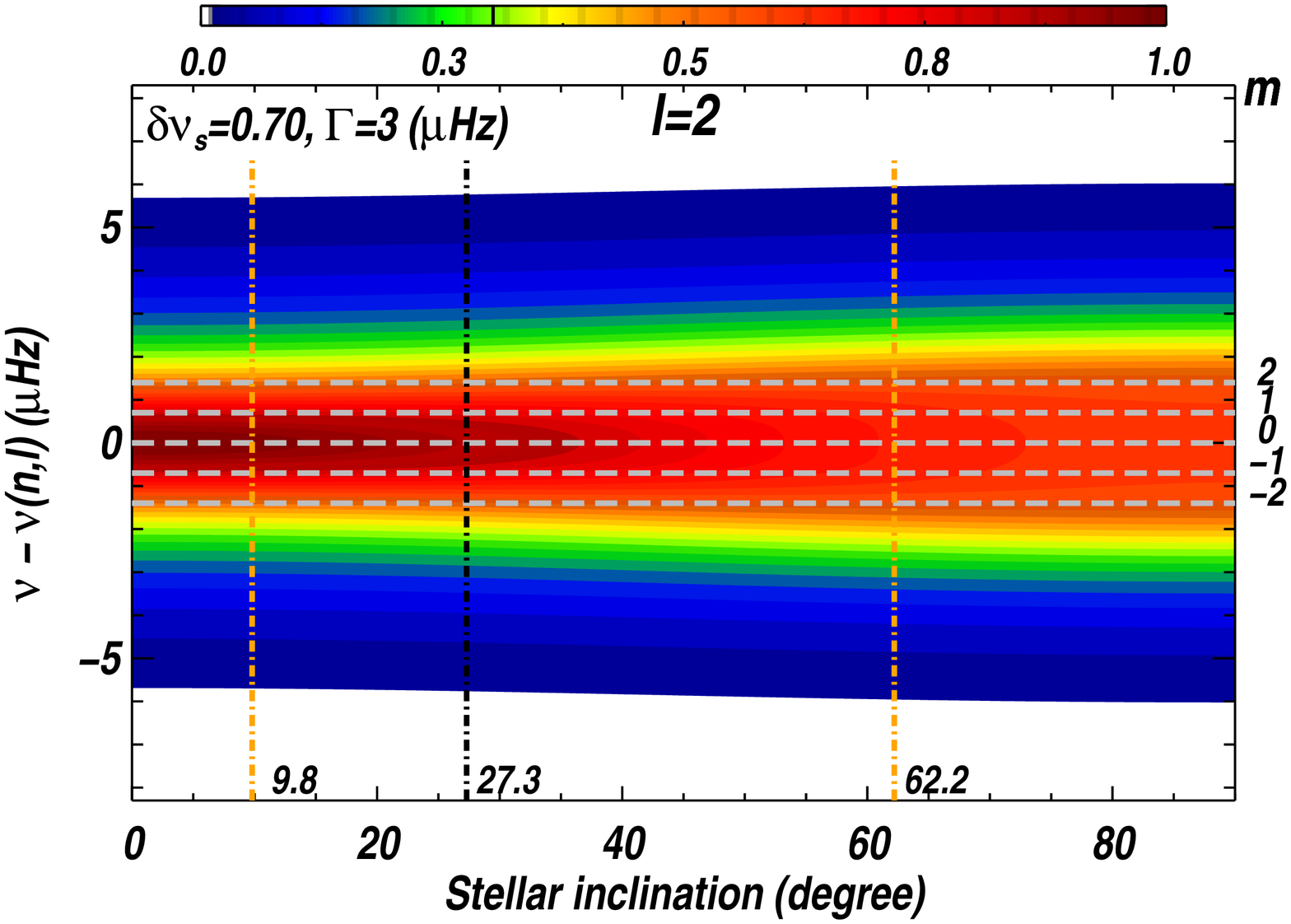, angle=90, width=8cm, height=6cm}}
  \end{center}
\caption{\textbf{HAT-P-7.} Relative power $P(\nu)$ of
azimuthal components for $l=1$ (three Lorentzian) and $l=2$ (five Lorentzian) at the median value of the
rotational splitting. Width of the Lorentzian is the average mode width
($\Gamma=3 \,\mu$Hz).  Horizontal dash lines indicate frequencies of the
multiplets $m$. Vertical dot-dash lines indicate the credible interval
(orange) and the median (black) of the measured inclination (cf. Figure
\ref{fig:is-pdfs-HATP7}). These figures show that because
$\delta\nu_{\rm s} \ll \Gamma$, the profile of modes of degree $l=1$ and
$l=2$ are almost insensitive to stellar inclination.}
\label{fig:is-nus-hatp7}
\end{figure*}

Figure \ref{fig:is-pdfs-HATP7} shows the joint probability density
function (PDF) of $\delta\nu_{\rm s}$ and $i_\star$,
$p(i_\star,\delta\nu_{\rm s})$, for HAT-P-7 as well
as their marginalised posterior PDF, $p(\delta\nu_{\rm s})$ and
$p(i_\star)$. As clearly illustrated, $i_\star$ of HAT-P-7 is not
tightly constrained. The most probable value is
$i_\star=\timeform{27D.3}{+\timeform{34D.9}\atop{-\timeform{17D.5}}}$
with a 68\% confidence interval. This suggests that the star is more
likely seen by its pole than by its equator, albeit with large
uncertainty. To understand why $i_\star$ is not well determined, we
show in Figure \ref{fig:is-nus-hatp7} the power $P(\nu)$ corresponding
to the modes of degree $l=1$ and $l=2$ as we did in Figure
\ref{fig:is-nus-ideal}, but we set the rotational splitting equal to the
observed median splitting ($\delta\nu_{\rm s} = 0.70\,\mu$Hz). The
width $\Gamma$ of each Lorentzian is fixed to the average width ($\Gamma
= 3$ $\mu$Hz) of the modes of the highest signal-to-noise ratio. In this
case, $\delta\nu_{\rm s} \ll \Gamma$ and the $m$ components cannot be
resolved. Thus, the mode profiles are almost insensitive to the stellar
inclination, contrary to the ideal case of well resolved modes as
illustrated in Figure \ref{fig:is-nus-ideal}.


Although $\delta\nu_{\rm s}$ is related to the average internal rotation
frequency\footnote{Each mode is sensitive to the rotation at a given
depth. Assuming a modest differential rotation, for low-degree p-modes,
$\delta\nu_{\rm s}$ is nearly equal to the surface rotation
frequency.}, it provides a good proxy to the surface rotation
frequency. Based on this idea, with the radius $R_\star$ derived by
stellar modelling, we calculated the seismic $v\sin i_\star = 2 \pi
R_\star\, \delta\nu_{\rm s} \sin i_\star$ (cf. Figure
\ref{fig:is-pdfs-HATP7}). We obtained $v\sin i_\star =
2.95^{+1.71}_{-1.98}$\,km\,s$^{-1}$, which is in agreement with $v\sin
i_\star = 3.8 \pm 0.5$\,km\,s$^{-1}$ obtained by P08.

The degeneracy in solutions due to the correlation between rotation and
inclination limits the precision. In our effort to improve our
constraint on the inclination angle, $i_\star$, we looked for signs of
surface rotation by computing the autocorrelation of the
timeseries. Solar-like stars may have long-lived surface stellar spots
at low latitude that can modulate the light flux periodically, thus
revealing the surface rotation period. Unfortunately, HAT-P-7 shows no
sign of activity. While this may indicate that the star is not
active, this is consistent with our interpretation of the small
inclination angle.

\section{Asteroseismology of Kepler-25 
\label{sec:asteroseismology:K25}}

Kepler-25 is an F star that shows oscillations for which we detected 30
modes of degree $l=0$, $1$ and $l=2$ spanning over 10 radial orders but
with amplitudes smaller than HAT-P-7 (Figure \ref{fig:spectrum:K25}).
	
\subsection{Fundamental Properties}
The precision on the extracted seismic frequencies is lower by
approximately a factor two, compared with the case of HAT-P-7. As seen
in the \'echelle diagram (Figure\,\ref{fig:EDs-K25}) the range of
observed frequencies does not allow us to entirely retrieve the
oscillation pattern of the modes, which certainly reduces the accuracy
of the modelling.

A seismic analysis of Kepler-25 has already been carried out by
\cite{Huber2013b} using the empirical scaling relations among
mass, radius, effective temperature, the frequency spacing
$\Delta\nu$ and frequency at maximum power of the modes, $\nu_{\rm max}$
[see for example \cite{Huber2011} for more details]. They derived
$M_\star=1.19 \pm 0.06\,M_{\odot}$ and $R_\star=1.309 \pm
0.023\,R_{\odot}$.

For this star, the model with the minimum $\chi^2$ is found with
surface effect and with an exponent of $b=4.9$.  It describes a star
with $M_\star=1.26 \pm 0.03\,M_{\odot}$ and $R_\star =1.34 \pm
0.01\,R_\odot$. This is consistent with the first estimates
by \cite{Huber2011}. The central hydrogen abundance of $X_{\rm c} =
0.329$ corresponds to 46.9\% of the initial hydrogen abundance,
suggesting a star in the middle of its main sequence stage.  The star
has a small convective core, extending up to $7\%$ of the stellar radius
and an outer convective zone representing $20\%$ of the stellar radius.
 
\subsection{Rotation and Inclination}

\begin{figure*}[t]
  \begin{center}
\subfigure{\epsfig{figure=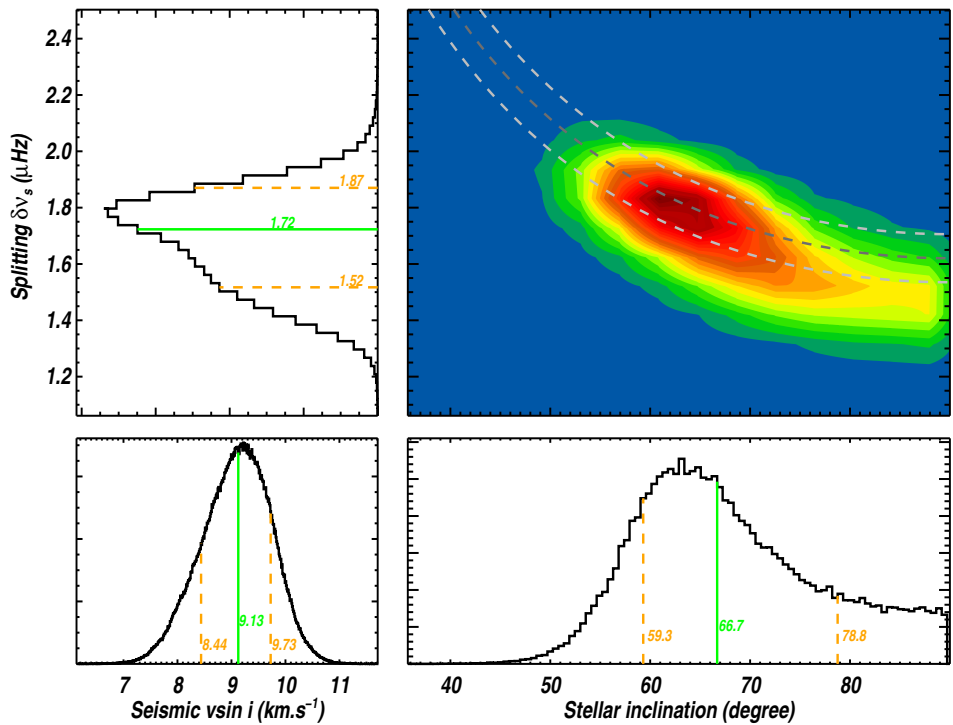, angle=90, width=16cm, height=12cm}}
  \end{center}
\caption{\textbf{Upper right.} Joint posterior probability distribution
of the stellar inclination and the rotation for Kepler-25. Blue
represents region of lowest probability. Red areas are of highest
probability. Superimposed and using a dark grey dotted line, we show the
spectroscopic $v\sin i_\star$ quoted by \citet{2014ApJS..210...20M} with
its $1\sigma$ uncertainty intervals (grey dotted lines). \textbf{Upper
left.} Marginalised probability density function for the rotational
splitting. \textbf{Lower right.} Marginalised probability density
function for the stellar inclination.  \textbf{Lower left.} Seismic
$v\sin i_\star$, inferred using the probability density for the
rotational splitting, the inclination and the radius of Kepler-25.}
\label{fig:is-pdfs-K25}
\end{figure*}

\begin{figure*}[t]
  \begin{center}
	\subfigure{\epsfig{figure=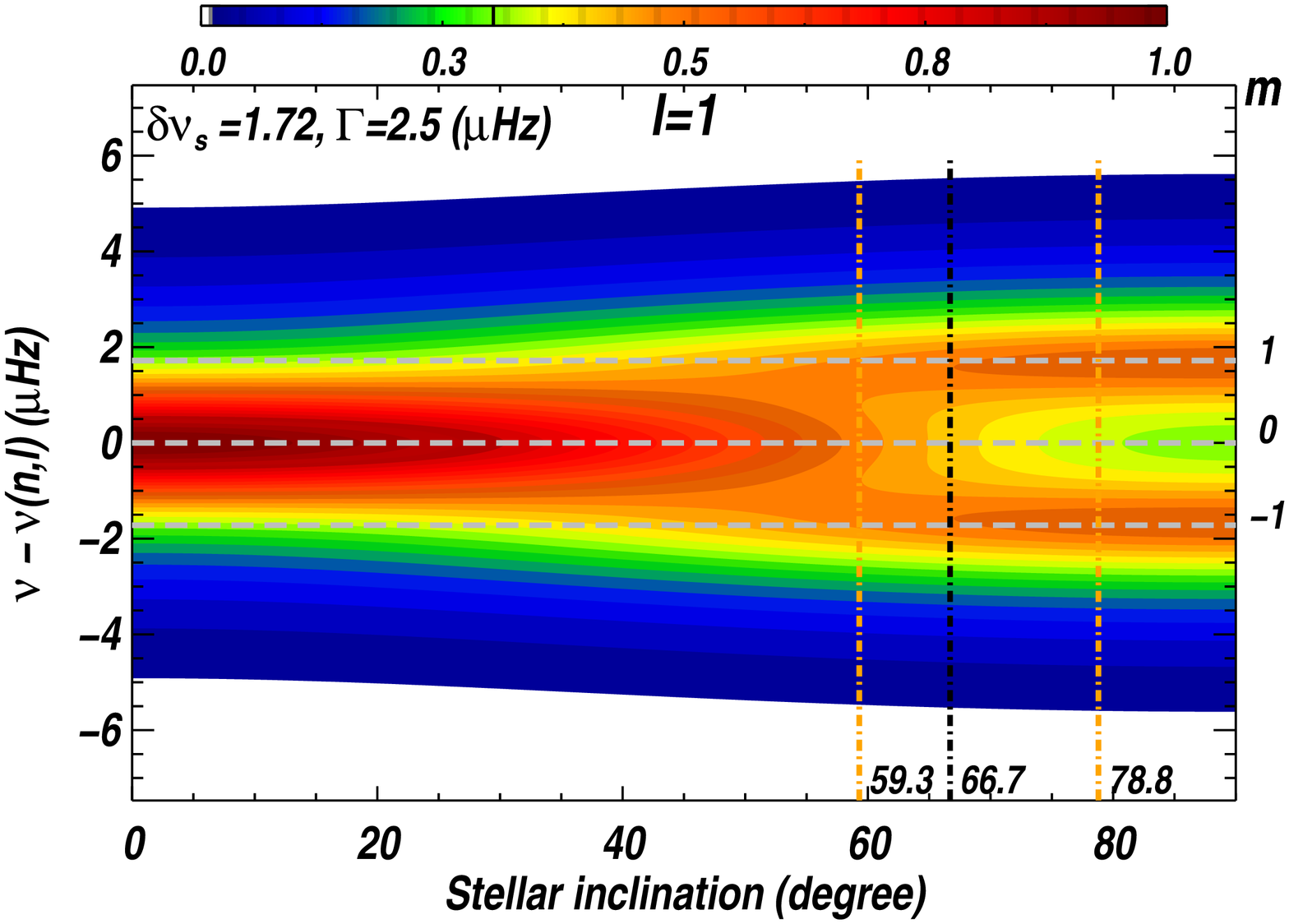, angle=90, width=8cm, height=6cm}}
	\subfigure{\epsfig{figure=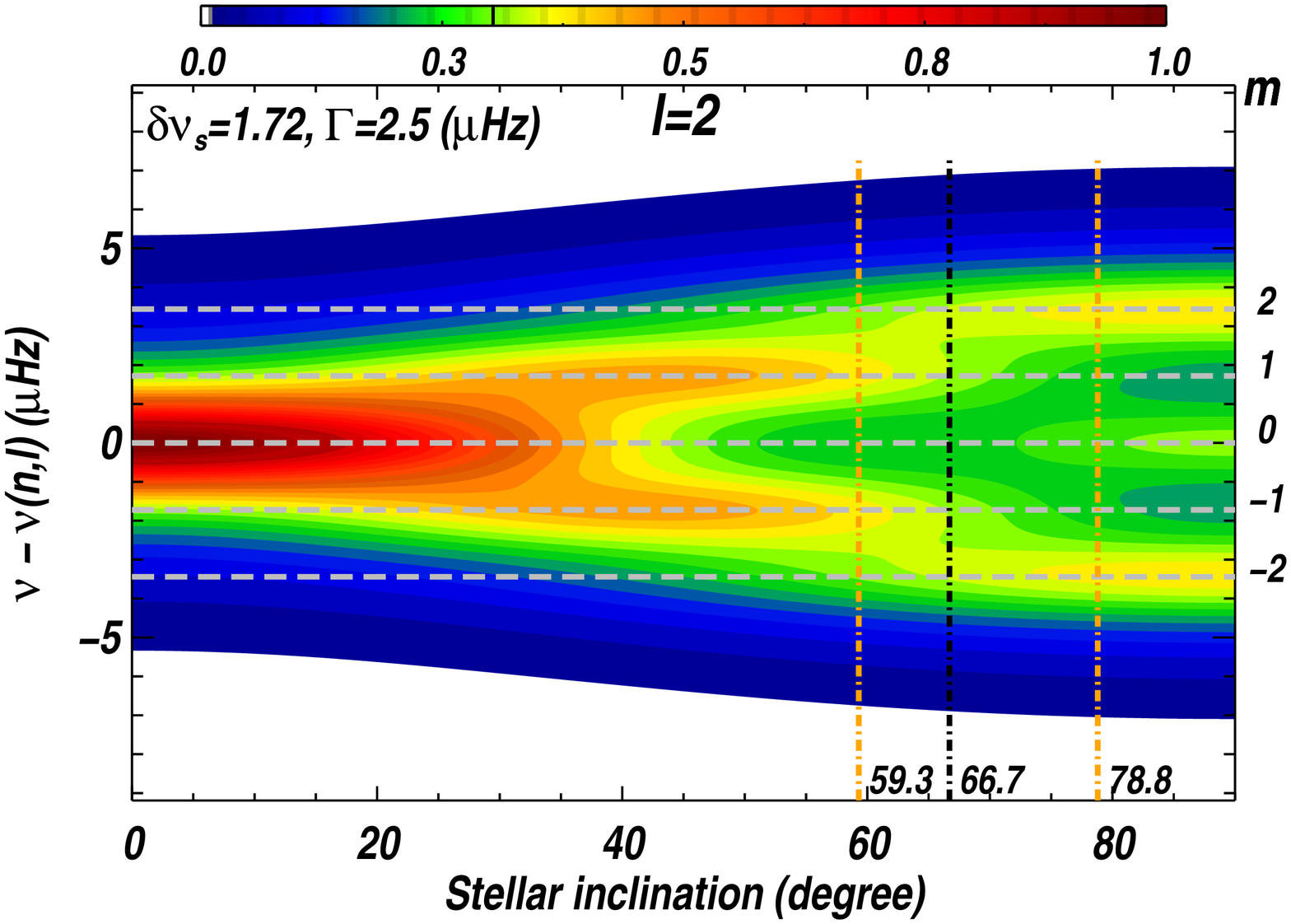, angle=90, width=8cm, height=6cm}}
  \end{center}
\caption{\textbf{Kepler-25.} Relative power {$P(\nu)$} of
azimuthal components for $l=1$ {(three Lorentzian)} and $l=2$ {(five Lorentzian)} at the median value of the
rotational splitting. Width of the Lorentzian is the average mode width
($\Gamma=2.5 \,\mu$Hz).  Horizontal dash lines indicate frequencies of
the multiplets $m$. Vertical dotted lines indicate the credible interval
(orange) and the median (black) of the measured inclination (cf. Figure
\ref{fig:is-pdfs-K25}). The rotation is fast enough to distinguish the
Lorentzian profiles of each azimuthal order. This allows an accurate
determination of the stellar inclination.}  \label{fig:is-nus-K25}
\end{figure*}

Figure \ref{fig:is-pdfs-K25} plots $p(i_\star,\delta\nu_{\rm s})$,
for Kepler-25 as well as their marginalised posterior PDF,
$p(\delta\nu_{\rm s})$ and $p(i_\star)$.  We obtain
$i_\star=\timeform{66D.7}{+\timeform{12D.1}\atop{-\timeform{7D.4}}}$
within a $68\%$ confidence interval.  The precision on $i_\star$
is much higher than for HAT-P-7, despite a lower signal-to-noise ratio.
This is because the rotational splitting is at least twice greater
($\delta\nu_s \simeq 1.72\,\mathrm{\mu Hz}$). The multiplets of each degree are
disentangled ($\delta\nu_s \approx \Gamma \simeq 2.5 \mu$Hz), and the
mode profile $\mathcal{E}(l,m, i_\star)$ becomes very sensitive to the
stellar inclination (cf. Figure \ref{fig:is-nus-K25}).

The radius derived from the best-fit model allows us to directly compare
the spectroscopically determined radial velocity, $v\sin i_\star$,
quoted by \citet{2014ApJS..210...20M} against our value. As shown in
Figure\,\ref{fig:is-pdfs-K25}, the spectroscopic $v\sin i_\star$ is
consistent with the maximum location of the joint
PDF. Moreover, the rotational kernels of the $l=1$ and $l=2$ modes
show that the measured rotational splitting is as much sensitive to the
rotation in the convective envelope as into the radiative zone.  The
modes are however not sensitive to the rotation in the inner convective
region. This indicates that the radiative layer and the outer convective
region are rotating uniformly, with the same velocity as the surface.
Finally, note that $H(n,l,m)$ autocorrelation of the timeseries does not show
evidence for stellar activity.

\section{Joint Analysis of the HAT-P-7 System
\label{sec:hat-p-7}}

{In this section and the next, we combine $i_\star$ from asteroseismology and $\lambda$ from the RM effect
to constrain the three-dimensional spin--orbit angle $\psi$.
Since the seismic $v\sin i_\star$ and $\rho_\star$ are also complementary to those from the RM effect and transit photometry,
we reanalyze the RM effect and the whole available {\it Kepler} lightcurves simultaneously, 
incorporating the constraints on $i_\star$, $v\sin i_\star$, and $\rho_\star$ described in the previous sections as the prior knowledge.
The method and results are presented in this section for HAT-P-7 and in the next section for Kepler-25.}

For the HAT-P-7 system, the combination of the asteroseismology and {\it Kepler} lightcurves provides 
a unique opportunity to tightly constrain the orbital eccentricity of HAT-P-7b,
especially because the occultation (secondary eclipse) is clearly detected for this giant and close-in planet.
Therefore, we first describe how the transit and {occultation} lightcurves constrain the planetary orbit 
in Section \ref{ssec:hatp7-lc}, before reporting the joint analysis for $\psi$ in Section \ref{ssec:hatp7-joint}.

\subsection{Analysis of Transit and Occultation Lightcurves
\label{ssec:hatp7-lc}}
\subsubsection{Data Processing and Revised Ephemeris}

In the following analysis, we use the {\it Kepler} short-cadence
Pre-search Data Conditioned Simple Aperture Photometry (PDCSAP) fluxes
through Q0 to Q17 retrieved from the NASA exoplanet
archive.\footnote{\texttt{http://exoplanetarchive.ipac.caltech.edu}}

First, lightcurves are detrended and normalized by fitting a third-order
polynomial to the out-of-transit fluxes around $\pm 0.5$ days of every
transit center.  Here, the central time and the duration of each transit
are determined from the central time of the first observed transit
calculated from the linear ephemeris, $t_0$, the orbital period, $P$,
and the duration taken from the archive.  We iterate the polynomial fit
until all the {$>5\sigma$} outliers are excluded.  In this process, we
remove the transits whose baselines cannot be determined reliably due to
the data gap around the ingress or egress.

Second, we fit each detrended and normalized transit with the lightcurve
model by \citet{2009ApJ...690....1O} to determine its central time.  We
fix the planet-to-star radius ratio, $R_{\rm p}/R_\star$, the ratio of
the semi-major axis to the stellar radius, $a/R_\star$, the cosine of
the orbital inclination, $\cos i_{\rm orb}$ {at those values} from
the archive, adopt the coefficients for the quadratic limb-darkening
law, $u_1$ and $u_2$, from \citet{2012ApJ...751..112J}, and assume zero
orbital eccentricity ($e$).  Since only the out-of-transit outliers were
removed in the first step, we also iteratively remove in-transit
{$>5\sigma$} outliers.  The resulting transit times are used to
phase fold all the transits and to improve the transit parameters and
orbital period $P$.

Using these revised transit parameters, we again fit each transit
lightcurve for its central time and total duration.  Here we assume
$e=0$, fix the values of $u_1$, $u_2$, $a/R_\star$, $R_{\rm p}/R_\star$,
and $P$, and float only central transit time and $\cos i_{\rm orb}$.
From these transit times, we calculate the revised ephemeris
$t_0(\mathrm{BJD}) - 2454833 = 121.3585049(49)$ and $P =
2.204735427(13)$ days by linear regression.  Since we find no systematic
TTVs, hereafter we assume that the orbit of HAT-P-7b is described by the
strictly periodic Keplerian orbit with $t_0$ and $P$ obtained above.

\subsubsection{Orbital Eccentricity and Mean Stellar Density from the Phase-folded Transit and Occultation\label{sssec:hatp7-lc-e}}

\begin{figure}
	\centering
\includegraphics[width=8.2cm,clip]{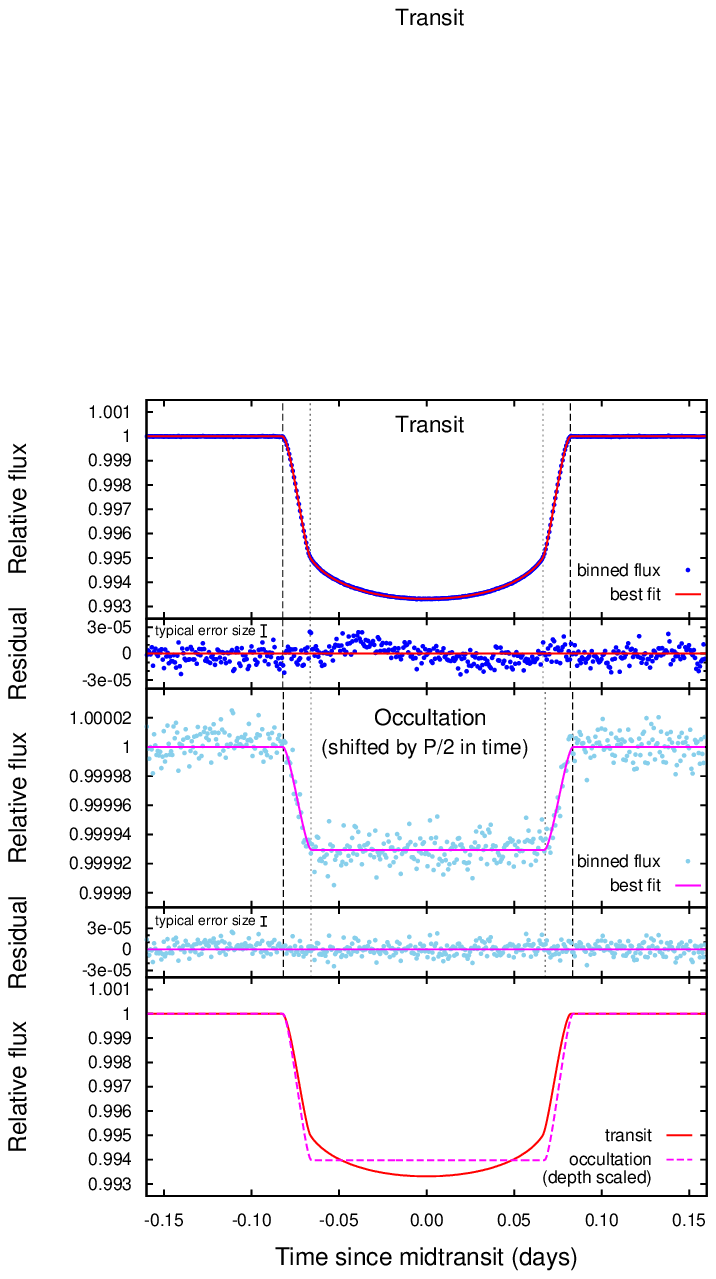}
\caption{Phase-folded transit (top) and occultation (middle)
	lightcurves.  Points are the binned fluxes ($1\,\mathrm{min}$)
	and solid lines show the best-fit model lightcurves.  Vertical
	dashed and dotted lines correspond to the four contact points;
	see figure 2 of \citet{2011exop.book...55W} for their
	definitions.  In the bottom panel, we compare the durations and
	central times of best-fit transit and occultation lightcurves.
	Occulation is shifted by $P/2$ in time in the middle and the
	bottom panels, and its depth is scaled by $\delta$ in the bottom
	panel for ease of comparison.}  \label{best_tra_occ}
\end{figure}

The top and middle panels of Figure \ref{best_tra_occ} respectively show
the transit and occultation lightcurves stacked using the revised
ephemeris.  The lightcurves are binned {into $1$-minute bins} and the
uncertainty of {the flux at the $i$-th bin},
$\sigma_{i,\mathrm{MAD}}$, is calculated as $1.4826 \times
\mathrm{median\ absolute\ deviation}$ divided by the square root of the
number of data points in the bin \citep{1969drea.book.....B}.  Solid
lines are the best-fit lightcurves obtained from the simultaneous fit to
both lightcurves.  We use the transit model by
\citet{2002ApJ...580L.171M}, and binned model fluxes are calculated by
averaging fluxes sampled at 0.1-minute interval.  { In this figure,
the transit and occultation are shifted in time by $t_{{\rm c,\,tra}}$
and $P/2+ t_{{\rm c,\,tra}}$, respectively, where $t_{{\rm c,\,tra}}$ is
the central time of the phase-folded transit lightcurve.  This parameter
is introduced to take into account the uncertainty in $t_0$, and the
best-fit value of $t_{{\rm c,\,tra}}$ is indeed within that uncertainty
(see Table \ref{tab:hatp7-joint}).}  In the transit residuals (top
panel), we reproduce the anomaly first reported by
\citet{2013ApJ...764L..22M}, who attributed it to the planet-induced
gravity darkening.

Since the asymmetry of the planetary orbit alters the relative duration
of the transit and occultation, as well as their time interval, one can
tightly constrain the orbital eccentricity from the combination of
transits and occultations; see equations (33) and (34) in
\citet{2011exop.book...55W} for instance.  {The bottom panel of
Figure \ref{best_tra_occ} illustrates this subtle effect by comparing
the best-fit transit and occultation lightcurves.  Here the depth of the
occultation is scaled by $\delta$, the occultation depth divided by $(R_{\rm
p}/R_\star)^2$, for ease of comparison.  In this panel, the
egress of the occultation occurs slightly later than that of the
transit, while the difference is smaller for their ingresses.  In other
words, our best-fit model indicates that the occultation duration is
longer than the transit one and that the center of occultation deviates
from $P/2$.}  These are most likely due to the asymmetry of the orbit
introduced by the slight but non-zero eccentricity, as well as the time
delay of $4.5\times10^{-4}$ days due to the finite speed of light (twice
the orbital semi-major axis divided by the speed of light; calculated
for $M_\star = 1.59\,M_\odot$).  In fact, with the non-zero eccentricity
and the above light-travel time included, the simultaneous fit to the
phase-folded transit and occultation lightcurves give tight constraints
on the planet's eccentricity, $e\cos\omega = 0.00026\pm0.00015$ and
$e\sin\omega = 0.0041\pm0.0022$, where $\omega$ is the argument of
periastron measured from the plane of the sky.

Since $e\sin\omega$ and $a/R_\star$ are degenerate in determining the
transit durations, the tight constraint on $e\sin\omega$ also allows the
accurate determination of $a/R_\star$, and hence the mean stellar
density $\rho_\star$ independently from asteroseismology
\citep{2003ApJ...585.1038S}.  We obtain $a/R_\star=4.131\pm0.009$ from
the above fit, and then deduce $\rho_\star = (0.275\pm0.002)\times
10^3\,\mathrm{kg\,m}^{-3}$ from
\begin{equation}
	\rho_\star = \frac{3\pi}{GP^2} \left(\frac{a}{R_\star}\right)^3 \left(1+\frac{M_{\rm p}}{M_\star}\right)^{-1} ,
\end{equation}
where $G$ denotes the gravitational constant, and $M_{\rm p}/M_\star
\sim 10^{-3}$ can be neglected.  This value is larger than
$\rho_{\star,\rm s}$ based on the seismic scaling relation by
$2.4\sigma$, but consistent with $\rho_{\star,\rm m}$ from the stellar
model at the $1\sigma$ level (see Table \ref{tab:output_model}).  For
this reason, we adopt the constraints from the stellar model as the
prior information in the following joint fit. The choice of the prior,
however, does not affect the spin--orbit angle determination, but only
slightly changes the values of $a/R_\star$, $\rho_\star$, $\cos i_{\rm
orb}$, and $e\sin\omega$.  The slight discrepancy between $\rho_\star$
from the seismic scaling relation ($\rho_{\star,\rm s}$) and that from
transit and occultation implies that the current precision of the {\it
Kepler} photometry has reached the level that could permit an
independent test of the seismic scaling relation for the mean stellar
density.

\subsection{Joint Analysis
\label{ssec:hatp7-joint}}
\subsubsection{Method
\label{sssec:hatp7-joint-method}}

In this subsection, we report the joint MCMC analysis of phase-folded
transit and occultation lightcurves (cf. Section \ref{ssec:hatp7-lc})
and RVs (cf. Section \ref{subsec:previous-hatp7}) making use of the
prior constraints on the mean stellar density $\rho_\star$, projected
stellar rotational velocity $v\sin i_\star$, and stellar inclination
$i_\star$ obtained from asteroseismology in Sections
\ref{sec:asteroseismology}--\ref{sec:asteroseismology:K25}.  As
discussed in Section \ref{ssec:hatp7-lc}, the precise constraint on
$\rho_\star$ (equivalent to that on $a/R_\star$) helps to lift the
degeneracy between $a/R_\star$ and $e\sin\omega$, thus resulting in
{improved constraints} on these {two} parameters.  In
addition, $v\sin i_\star$ is the key parameter for the RM effect along
with $\lambda$, and so the constraints on $v\sin i_\star$ help us to
better determine $\lambda$ from the observed RM signal.  Finally,
$i_\star$ is crucial in determining the three-dimensional spin--orbit
angle $\psi$ via Equation (\ref{psi_3d}), which is the {major goal}
of this paper.

In order to properly handle the possible correlation among $\lambda$,
$v\sin i_\star$, and $i_\star$, we adopt the joint probability
distribution for $v\sin i_\star$ and $i_\star$ as the prior in our MCMC
analysis and directly calculate the posterior distribution for $\psi$ by
floating $i_\star$ as well.
It should be noted here that our observables {do not determine} the
sign of $\cos i_\star$ or $\cos i_{\rm orb}$, due to the symmetry with
respect to the plane of the sky.  In order to take into account this
inherent degeneracy, we randomly change the sign of the first term in
Equation (\ref{psi_3d}) in computing $\psi$.  Since the probability
distribution of $\rho_\star$ is almost independent {of} those of
$v\sin i_\star$ and $i_\star$, we include the constraint on this
parameter as an independent Gaussian with the central value and width of
$\rho_{\star,\rm m}$ {listed} in Table \ref{tab:output_model}.

We adopt the same model (including non-zero eccentricity and
light-travel time) for transit and occultation as in Section
\ref{ssec:hatp7-lc}.  {The observed} RVs are modeled as
\begin{equation}
	v_{\star, \mathrm{model}}(t) = v_{\star,\mathrm{orb}}(t) + v_{\star, \mathrm{RM}}(t) + \gamma_i + \dot \gamma (t - t_0).
\end{equation}
Here, 
\begin{equation}
	v_{\star,\mathrm{orb}} = K_\star \left[ \cos(\omega + f) + e \cos \omega \right]
\end{equation}
is the stellar orbital RVs for the Keplerian orbit, where $K_\star$ is
the RV semi-amplitude and $f$ is the true anomaly of the planet.  The
$\gamma_i$ ($i=1, 2$) are the constant offsets for RVs from Keck/HIRES
($i=1$) and Subaru/HDS ($i=2$), and $\dot{\gamma}$ accounts for the
linear trend in the observed RVs in the W09 data set \citep{Winn2009,
2012PASJ...64L...7N, 2014ApJ...785..126K}.  Finally, anomalous RVs due
to the RM effect, $v_{\star, \mathrm{RM}}$, are modeled following
\citet{Hirano2011}.  The parameters {characterizing} the RM model
{include} $v\sin i_\star$ (projected rotational velocity of the
star), $\beta$ (Gaussian dispersion of spectral lines), $\gamma$
(Lorentzian dispersion of spectral lines), $\zeta$ (macroturbulence
dispersion of spectral lines), $u_{1\mathrm{RM}}+u_{2\mathrm{RM}}$, and
$u_{1\mathrm{RM}}-u_{2\mathrm{RM}}$ (coefficients for the quadratic
limb-darkening law in the RM effect).  We do not take into account the
effect of convective blueshift \citep{2011ApJ...733...30S}, as its
typical amplitude ($\sim 1\,\mathrm{m\,s^{-1}}$) is smaller than the
(jitter-included) precision of {the} RVs analyzed here.

We impose the non-seismic priors as well {on} some of the model
parameters.  For the ephemeris, we use the Gaussian priors $t_0
(\mathrm{BJD}) -2454833 = 121.3585049 \pm 0.0000049$ and $P =
2.204735427 \pm 0.000000013\,\mathrm{days}$ obtained from the transit
lightcurves.  The priors on the RM parameters {($\beta$, $\gamma$, $\zeta$,
$u_{1\rm RM} + u_{2\rm RM}$, and $u_{1\rm RM} - u_{2\rm RM}$)} are almost the same as in
A12. Namely, we fix $\beta=3\,\mathrm{km\,s^{-1}}$ and
$\gamma=1\,\mathrm{km\,s^{-1}}$, and assume Gaussian prior $\zeta = 5.18
\pm 1.5\,\mathrm{km\,s^{-1}}$.  We fix the value of
$u_{1\mathrm{RM}}-u_{2\mathrm{RM}}$ at $-0.023$ from the tables of
\citet{2000A&A...363.1081C} for the Johnson V band and the ATLAS model.
The value is obtained using the {\tt jktld}
tool\footnote{\texttt{http://www.astro.keele.ac.uk/jkt/codes/jktld.html}}
for the parameters $T_{\rm eff}=6350\,\mathrm{K}$, $\log
g\,(\mathrm{cgs})=4.07$, and $[{\rm Fe/H}]=0.3$. The value of
$u_{1\mathrm{RM}}+u_{2\mathrm{RM}}$ is floated around the tabulated
value of $0.70$ assuming the Gaussian prior of width $0.10$.  In
addition, we impose an additional Gaussian prior on $v\sin i_\star$
based on the spectroscopic value in Table \ref{tab:nonseism_obs},
because the seismic constraint on this parameter is independent {of}
the spectroscopic $v\sin i_\star$.  We assume uniform priors for the
other $13$ fitting parameters {listed} in Table
\ref{tab:hatp7-joint} (top and middle blocks).

In the joint fit, we assume the same values of stellar jitter as used in
the original papers{;} $9.3\,\mathrm{m\,s^{-1}}$ for the W09 set,
$3.8\,\mathrm{m\,s^{-1}}$ for the Keck/HIRES RVs of the N09 set, and
$6.0\,\mathrm{m\,s^{-1}}$ for the A12 set.  In order to prevent the
transit and occultation lightcurves from placing unreasonably tight
constraints compared to RVs, we also {increase the errors quoted
for} photometric data (evaluated in Section \ref{sssec:hatp7-lc-e}) as
$\sigma_i = \sqrt{ \sigma_{i,\mathrm{MAD}}^2 + \sigma_{\rm r}^2}$.
Here, $\sigma_{\rm r} = 5.8 \times 10^{-6}$ is a parameter analogous to
the RV jitter and chosen so that the reduced $\chi^2$ of the lightcurve
fit becomes unity.  This prescription is also motivated by the following
two facts.  First, $\sigma_{i,\mathrm{MAD}}$ tends to underestimate the
true uncertainty because it neglects the effect of correlated noise.
Indeed, when the number of data points is sufficiently large,
uncertainties are dominated by the correlated or ``red'' noise component
\citep{2006MNRAS.373..231P}.  Second, the systematic residuals of the
best-fit transit model (top panel of Figure \ref{best_tra_occ}) suggest
other effects that are not taken into account in our model [e.g.,
possible planet-induced gravity darkening discussed by
\citet{2013ApJ...764L..22M}].  Placing too much weights on such features
could bias the transit parameters.\\

\subsubsection{Results}

Constraints on the system parameters from the joint analysis are
 summarized in Table \ref{tab:hatp7-joint}.  The ``parameters mainly
 derived from lightcurves/RVs'' are the model (fitted) parameters, while
 the ``derived quantities'' are the parameters derived from the model
 parameters (along with $M_\star$ and $R_\star$ in Table
 \ref{tab:output_model} for $M_{\rm p}$, $R_{\rm p}$, and $\rho_{\rm
 p}$).  While our result {is in a reasonable agreement} with previous
 studies \citep[c.f.,][]{2013ApJ...764L..22M, 2013ApJ...772...51E,
 2013ApJ...774L..19V}, it provides two major improvements.

First, we determine the orbital eccentricity of HAT-P-7b essentially
from the photometry (i.e., transit, occultation, and asteroseismology)
alone.  A similar method has recently been {employed} by
\citet{2014ApJ...782...14V} to constrain the planet's orbital
eccentricity using the seismic stellar density \citep[see
also][]{2012ApJ...756..122D, 2014MNRAS.440.2164K}, but here we show that
this method is {also useful} for such a low-eccentricity orbit.
Furthermore, our result is even more precise and reliable because
{it takes into account} the independent constraint on $\rho_\star$ and $e$
from the occultation lightcurve.

\begin{figure}
	\centering \includegraphics[bb= 50 60 200 300,
	width=8.5cm]{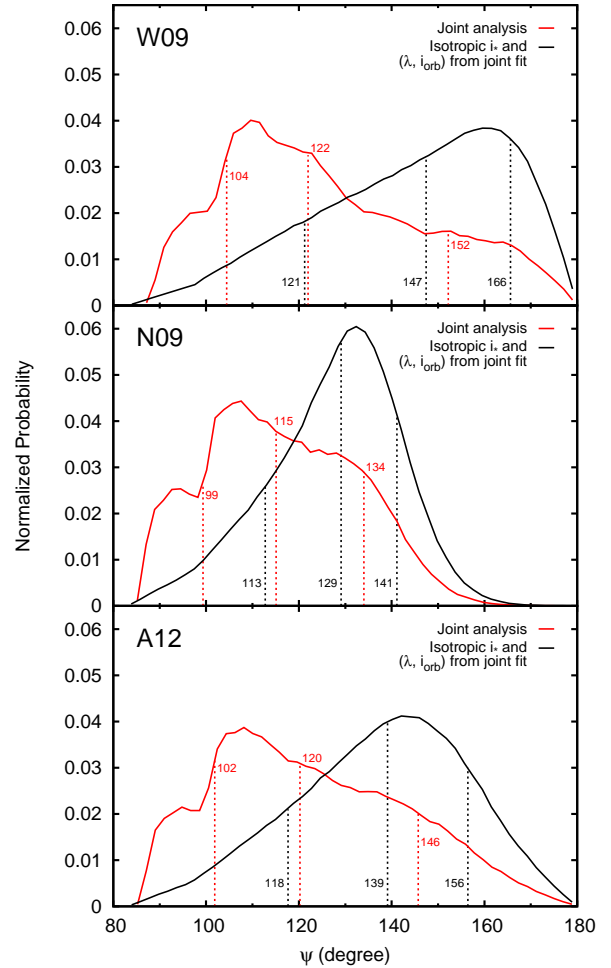} 
\caption{Probability distributions for the three-dimensional spin--orbit
	angle $\psi$ of HAT-P-7b for the W09 (top), N09 (middle), and
	A12 (bottom) data sets.  Solid red lines show the posteriors
	from the joint analysis, while the black ones are the
	probability distributions obtained from uniform $\cos i_\star$
	and the posteriors of $\lambda$ and $i_{\rm orb}$ from the joint
	analysis (Table \ref{tab:hatp7-joint}).  The median, $1\sigma$
	lower limit, and $1\sigma$ upper limit for each distribution are
	shown in vertical dotted lines.  A small bump around $\psi
	\approx 95^\circ$ in each panel originates from the fact that
	each posterior shown here is the superposition of the two
	inherently degenerate configurations with the opposite signs of
	$\cos i_\star \cos i_{\rm orb}$; see the discussion in the
	second paragraph of Section \ref{sssec:hatp7-joint-method}.}
	\label{psi_hatp7}
\end{figure}

Second, we obtain the probability distribution for the three-dimensional
spin--orbit angle $\psi$ in a consistent manner.  In the case of
HAT-P-7, the constraint on $\psi$ is not very strong because the modest
splitting of the azimuthal modes only allows a weak constraint on
$i_\star$ (see Figure \ref{fig:is-nus-hatp7}).  Nevertheless, we find
that the peak values of $\psi$ shift towards $90^\circ$ compared to
those obtained from the ``random'' $i_\star$ (uniform in $\cos i_\star$)
in all three data sets, as shown in Figure \ref{psi_hatp7}.  Moreover,
the methodology presented here can be applied to other systems, for some
of which asteroseismology may be able to tightly constrain $i_\star$
unlike HAT-P-7.  We will show that this is indeed the case for the
Kepler-25 system in the next section.

\renewcommand{\arraystretch}{0.7} 
\begin{table*}
	\caption{Parameters of the HAT-P-7 System from the Joint Analysis.}	
	\centering
	\label{tab:hatp7-joint}
	\small
	\begin{tabular}{cccc}
	\toprule
	Parameter						& 	 Value (W09) 	& 	Value (N09)	&	Value (A12)\\
	\midrule	
	\multicolumn{4}{c}{\it Parameters mainly derived from lightcurves (transit, occultation, asteroseismology)}\\
	\midrule
	$t_0(\mathrm{BJD})-2454833$		& \multicolumn{3}{c}{$121.3585049 \pm 0.0000049$}\\
	$P$ 	(days)						& \multicolumn{3}{c}{$2.204735427 \pm 0.000000013$}\\
	$e\cos\omega$ 					& $0.00024\pm0.00020$ & $0.00024\pm0.00020$ & $0.00025\pm0.00020$\\	
	$e\sin\omega$ 					& $0.0053_{-0.0021}^{+0.0022}$ & $0.0057_{-0.0026}^{+0.0025}$ & $0.0049_{-0.0030}^{+0.0026}$\\
	$u_1$ 							& $0.3540\pm0.0034$ & $0.3544_{-0.0034}^{+0.0033}$ & $0.3545_{-0.0035}^{+0.0034}$\\
	$u_2$ 							& $0.1670_{-0.0054}^{+0.0055}$ & $0.1663_{-0.0053}^{+0.0055}$ & $0.1661_{-0.0055}^{+0.0056}$\\
	$\rho_\star$ ($10^3\,\mathrm{kg\,m^{-3}}$) & $0.2736\pm0.0016$ & $0.2731_{-0.0018}^{+0.0021}$ & $0.2737_{-0.0018}^{+0.0024}$\\
	$\cos i_{\rm orb}$ 			& $0.12149_{-0.00057}^{+0.00056}$ & $0.12166_{-0.00068}^{+0.00063}$ & $0.12145_{-0.00081}^{+0.00061}$\\
	$R_{\rm p}/R_\star$ 				& $0.077589_{-0.000021}^{+0.000020}$ & $0.077593\pm0.000020$ & $0.077591_{-0.000021}^{+0.000020}$\\
	$\delta$							& \multicolumn{3}{c}{$0.01171\pm0.00010$}\\		
	{$t_{{\rm c,\,tra}}$} (days)			& \multicolumn{3}{c}{$-0.0000044_{-0.0000042}^{+0.0000041}$}\\
	$i_\star$ ($^\circ$)					& $31_{-16}^{+33}$ 	& $33_{-20}^{+34}$	& $33_{-20}^{+34}$\\
	\midrule
	\multicolumn{4}{c}{\it Parameters mainly derived from RVs}\\
	\midrule
	$K_\star$ ($\mathrm{m\,s^{-1}}$)	& $211.7\pm2.3$	& $213.2\pm1.8$	& $214.0\pm4.6$\\	
	$\gamma_1$ ($\mathrm{m\,s^{-1}}$)	& $-15.5\pm3.0$ 	& $-37.5\pm1.5$	& $10.4_{-1.6}^{+1.5}$ \\
	$\gamma_2$ ($\mathrm{m\,s^{-1}}$)	& $-9.7\pm1.7$ 	& $-16.9\pm1.4$	& --\\
	$\dot{\gamma}$ ($\mathrm{m\,s^{-1}\,yr^{-1}}$)	& $21.5\pm2.5$ 	& -- & --\\
	$\lambda$ ($^\circ$)				& $186_{-11}^{+10}$ 	& $220.3_{-9.3}^{+8.2}$	& $157_{-13}^{+14}$\\
	$v\sin i_\star$ ($\mathrm{km\,s^{-1}}$)& $4.15_{-0.39}^{+0.38}$ & $3.17\pm0.33$ 		& $3.17_{-0.34}^{+0.33}$\\
	$\beta$ ($\mathrm{km\,s^{-1}}$)		&	\multicolumn{3}{c}{$3.0$ (fixed)}\\
	$\gamma$ ($\mathrm{km\,s^{-1}}$)	&	\multicolumn{3}{c}{$1.0$ (fixed)}\\
	$\zeta$ ($\mathrm{km\,s^{-1}}$)		& $5.3\pm1.5$	& $5.5\pm1.5$	& $5.5\pm1.5$\\
	$u_{1\mathrm{RM}}+u_{2\mathrm{RM}}$	& \multicolumn{3}{c}{$0.70\pm0.10$}\\
	$u_{1\mathrm{RM}}-u_{2\mathrm{RM}}$	&	\multicolumn{3}{c}{$-0.23$ (fixed)}\\
	\midrule
	\multicolumn{4}{c}{\it Derived quantities}\\
	\midrule
	$\psi$ ($^\circ$) 						& $122_{-18}^{+30}$	& $115_{-16}^{+19}$	& $120_{-18}^{+26}$\\
	$a/R_\star$ 						& $4.1269_{-0.0078}^{+0.0082}$	& $4.1245_{-0.0092}^{+0.0103}$ & $4.1277_{-0.0090}^{+0.0121}$\\
	transit impact parameter ($R_\star$) 	& $0.4987\pm0.0013$		 & $0.4989\pm0.0013$			& $0.4988_{-0.0014}^{+0.0013}$\\
	$T_{14,\mathrm{tra}}$ (days) 		& $0.164301\pm0.000022$	& $0.164303\pm0.000023$		& $0.164300\pm0.000023$\\
	$T_{23,\mathrm{tra}}$ (days) 	& $0.133042_{-0.000048}^{+0.000049}$ & $0.133034_{-0.000048}^{+0.000047}$& $0.133037_{-0.000048}^{+0.000052}$\\
	$T_{\mathrm{tra}}$ (days) 			& $0.148672_{-0.000024}^{+0.000025}$	& $0.148668\pm0.000024$	& $0.148669_{-0.000024}^{+0.000025}$\\
	occultation impact parameter ($R_\star$)  	& $0.5040_{-0.0023}^{+0.0022}$		& $0.5047_{-0.0028}^{+0.0025}$ & $0.5039_{-0.0033}^{+0.0024}$\\
	$T_{14,\mathrm{occ}}$ (days) 	& $0.16555_{-0.00050}^{+0.00051}$	& $0.16566_{-0.00061}^{+0.00058}$ & $0.16547_{-0.00070}^{+0.00060}$\\
	$T_{23,\mathrm{occ}}$ (days) 	& $0.13385_{-0.00033}^{+0.00034}$	& $0.13392_{-0.00040}^{+0.00039}$ & $0.13379_{-0.00046}^{+0.00041}$\\
	$T_{\mathrm{occ}}$ (days) 		& $0.14970_{-0.00041}^{+0.00042}$	& $0.14979_{-0.00051}^{+0.00048}$ & $0.14963_{-0.00058}^{+0.00050}$\\
	occultation depth (ppm)			& \multicolumn{3}{c}{$70.5\pm0.6$}\\
	$M_{\rm p} (M_{\rm J})$ 		& $1.86\pm0.03$ & $1.87\pm0.03$ & $1.88\pm0.05$\\
	$R_{\rm p} (R_{\rm J})$ 			& \multicolumn{3}{c}{$1.526\pm0.008$}\\
	$\rho_{\rm p}$ ($10^3\,\mathrm{kg\,m^{-3}}$)	& $0.65\pm0.01$ & $0.66\pm0.01$ & $0.66\pm0.02$\\
	\bottomrule
	\\
	\multicolumn{4}{l}{\parbox{0.9\textwidth}{ 
	Note --- The quoted best-fit values are the medians of their MCMC posteriors, 
	and uncertainties exclude 15.87\% of values at upper and lower extremes.
	The $T_{ij}$ ($i, j = 1, 2, 3, 4$) is the duration between the two contact points $i$ and $j$
	[see figure 2 of \citet{2011exop.book...55W} for their definitions], and $T=(T_{14} + T_{23})/2$.
	The subscript ``tra"  refers to transits and ``occ" to occultations.
	}
	}
	\end{tabular}
\end{table*}
\renewcommand{\arraystretch}{1.0} 

\section{Joint Analysis of the Kepler-25 System
\label{sec:kepler-25}}

\subsection{Method
\label{ssec:method-kepler-25}}

We repeat almost the same analysis for Kepler-25c as in Section
\ref{sec:hat-p-7}.  There are, however, several differences in the
lightcurve and RV analyses as described below, mainly due to the
multiplicity of the Kepler-25 system and relatively small
signal-to-noise ratio of the Kepler-25c's transit:
\begin{enumerate}
\item We phase-fold the transits using the actually observed transit
times rather than those calculated from the linear ephemeris.  This is
because the transit times of Kepler-25c ($P = 12.7$ days) exhibit
significant TTVs due to the proximity to the $2:1$ mean-motion resonance
with Kepler-25b ($P = 6.2$ days).  This is why we do not allow
{$t_{{\rm c, \,tra}}$}, the central time of the phase-folded
transit, to be a free parameter.  We adopt $\sigma_{\rm r} = 1.6 \times
10^{-5}$ based on the $\chi^2$ of the lightcurve fit.
\item The {occultation} of Kepler-25c was not detected and not taken
      into account in the following analysis.
\item As the quality of the transit lightcurve of Kepler-25c is not so
      good as that of HAT-P-7b, we could not determine the
      limb-darkening coefficients very well.
For this reason, we impose the prior $u_1-u_2=-0.0015\pm0.50$ based on the tables of \citet{2000A&A...363.1081C}, and choose 
{$u_1+u_2$ and $u_1-u_2$, instead of $u_1$ and $u_2$, as free
      parameters.}
We {made sure} that the choice of the confidence interval for
      $u_1-u_2$ does not affect the constraint on $\psi$.
\item In order to take into account the other planets in the RV fit, we
      allow the orbital semi-amplitude $K_\star$ and RV offset $\gamma$
      for each of the nights in 2011 and 2012 to be free parameters, as
      in A13. RV jitters are fixed at $3.3\,\mathrm{m\,s^{-1}}$.
\item We do not fit the orbital eccentricity but fix $e=0$, because we
      do not analyze the occultation nor RVs throughout the orbit
      \citep{2014ApJS..210...20M}.
\item We assume the independent Gaussian priors
      $u_{1\mathrm{RM}}+u_{2\mathrm{RM}} = 0.69\pm0.10$ and $\zeta =
      4.85\pm1.5\,\mathrm{km\,s^{-1}}$ from A13, and fix
      $u_{1\mathrm{RM}}-u_{2\mathrm{RM}} = -0.0297$ from the tables of
      \citet{2000A&A...363.1081C}.
\end{enumerate}

\subsection{Results
\label{ssec:results-kepler-25}}

\begin{figure}
	\centering
	\includegraphics[width=8.5cm,clip]{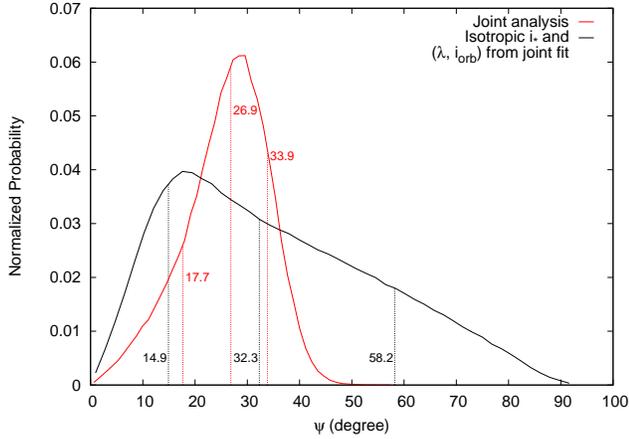}
	\caption{Probability distributions for the three-dimensional
	spin--orbit angle $\psi$ of Kepler-25c.  The solid red line
	shows the posterior from the joint analysis, while the black one
	is the probability distribution obtained from $\lambda$ and
	$i_{\rm orb}$ in Table \ref{tab:kepler25-joint} and uniform
	$\cos i_\star$.  The median, $1\sigma$ lower limit, and
	$1\sigma$ upper limit for each distribution are shown in
	vertical dotted lines.}  \label{psi_kep25}
\end{figure}

In the case of the Kepler-25 system, the uncertainty in $\psi$ is
significantly reduced by virtue of the seismic information.  This
situation is clearly illustrated in Figure \ref{psi_kep25}, which
compares the posterior probability distribution for $\psi$ from the
joint fit (solid red line) to that based on $\lambda$ and $i_{\rm orb}$
from the joint fit and the uniform $\cos i_\star$ (solid black line).
The corresponding system parameters are summarized in Table
\ref{tab:kepler25-joint}. They are basically consistent with
those {obtained} by A13, except for the increased precision in the
transit parameters.

Interestingly, our result suggests a spin--orbit misalignment for
Kepler-25c with more than $2\sigma$ significance.  In order to check the
robustness of this result, we also calculate the probability
distribution of $\psi$ for the seismic $i_\star$ and an independent
Gaussian {$\lambda=-\timeform{0D.5}\pm\timeform{5D.7}$} from the
Doppler tomography.  We obtain
$\psi=\timeform{23D.7}_{-\timeform{11D.3}}^{+\timeform{8D.0}}$ in this
case, which still {points to} the spin--orbit misalignment
marginally.  If confirmed, this will be the first example of the
spin--orbit misalignment in the multi-transiting system around a
main-sequence star\footnote{The first spin--orbit misalignment in the
multi-transiting system was confirmed by \citet{huber2013} around a red
giant star Kepler-56; they also used asteroseismology.}.  The
implication of this result will be discussed in Section
\ref{subsec:summary_K25}.

\begin{table}
	\caption{Parameters of the Kepler-25 System from the Joint Analysis (see also the note in Table \ref{tab:hatp7-joint}).}	
	\centering
	\label{tab:kepler25-joint}
	\begin{tabular}{cc}
	\toprule
	Parameter						& 	 Value (A13)\\

	\midrule	
	\multicolumn{2}{c}{\it Parameters mainly derived from lightcurves}\\
	\multicolumn{2}{c}{\it (transit, asteroseismology)}\\
	\midrule
	$t_0(\mathrm{BJD})-2454833$		& $127.646558_{-0.000094}^{+0.000096}$\\
	$P$ 	(days)					& $12.7203724_{-0.0000013}^{+0.0000014}$\\
	$u_1 + u_2$ 					& $0.550\pm0.018$\\
	$u_1 - u_2$ 					& $-0.27\pm0.44$\\
	$\rho_\star$ ($10^3\,\mathrm{kg\,m^{-3}}$) & $0.733_{-0.012}^{+0.013}$\\
	$\cos i_{\rm orb}$ 				& $0.04788_{-0.00038}^{+0.00036}$\\
	$R_{\rm p}/R_\star$ 				& $0.03590_{-0.00046}^{+0.00054}$\\
	$i_\star$ ($^\circ$)				& $65.4_{-6.4}^{+10.6}$ \\
	\midrule	
	\multicolumn{2}{c}{\it Parameters mainly derived from RVs}\\
	\midrule
	$K_{\star,2011}$ ($\mathrm{m\,s^{-1}}$)	& $-13\pm22$\\		
	$K_{\star,2012}$ ($\mathrm{m\,s^{-1}}$)	& $-37\pm30$\\	
	$\gamma_{2011}$ ($\mathrm{m\,s^{-1}}$)	& $-3.5\pm1.3$\\
	$\gamma_{2012}$ ($\mathrm{m\,s^{-1}}$)	& $2.0\pm1.4$ \\
	$\lambda$ ($^\circ$)					& $9.4 \pm 7.1$ \\
	$v\sin i_\star$ ($\mathrm{km\,s^{-1}}$)& $9.34_{-0.39}^{+0.37}$ \\
	$\beta$ ($\mathrm{km\,s^{-1}}$)		&	\multicolumn{1}{c}{$3.0$ (fixed)}\\
	$\gamma$ ($\mathrm{km\,s^{-1}}$)	&	\multicolumn{1}{c}{$1.0$ (fixed)}\\
	$\zeta$ ($\mathrm{km\,s^{-1}}$)		& $4.9\pm1.5$\\
	$u_{1\mathrm{RM}}+u_{2\mathrm{RM}}$	& $0.69\pm0.10$\\
	$u_{1\mathrm{RM}}-u_{2\mathrm{RM}}$	&	\multicolumn{1}{c}{$-0.0297$ (fixed)}\\
	\midrule	
	\multicolumn{2}{c}{\it Derived quantities}\\
	\midrule
	$\psi$ ($^\circ$) 					& $26.9_{-9.2}^{+7.0}$\\
	$a/R_\star$ 					& $18.44\pm0.11$\\
	transit impact parameter ($R_\star$)  					& $0.8826\pm0.0018$\\
	$T_{14,\mathrm{tra}}$ (days) 		& $0.11925\pm0.00025$\\
	$T_{23,\mathrm{tra}}$ (days) 		& $0.08528_{-0.00069}^{+0.00065}$\\
	$T_{\mathrm{tra}}$ (days) 			& $0.10226_{-0.00037}^{+0.00036}$\\
	\bottomrule
	\end{tabular}
\end{table}

\section{Discussion
\label{sec:summary}}

\subsection{HAT-P-7
\label{subsec:summary_hatp7}}

From asteroseismology alone, we obtain
$i_\star=\timeform{27D}_{-\timeform{18D}}^{+\timeform{35D}}$ for HAT-P-7
(Figure \ref{fig:is-pdfs-HATP7}).  This constraint, combined with the
{\it Kepler} lightcurves and the three independent RM measurements,
yields $\psi=\timeform{122D}_{\timeform{-18D}}^{\timeform{+30D}}$ and
$i_\star=\timeform{31D}^{+\timeform{33D}}_{-\timeform{16D}}$,
$\psi=\timeform{115D}_{\timeform{-16D}}^{\timeform{+19D}}$ and
$i_\star=\timeform{33D}^{+\timeform{34D}}_{-\timeform{20D}}$, and
$\psi=\timeform{120D}_{\timeform{-18D}}^{\timeform{+26D}}$ and
$i_\star=\timeform{33D}_{-\timeform{20D}}^{+\timeform{34D}}$ for the RVs
from W09, N09, and A12, respectively (Figure \ref{psi_hatp7} and Table
\ref{tab:hatp7-joint}).  Although the resulting constraints are not very
strong due to the modest splittings of azimuthal modes (see Figure
\ref{fig:is-nus-hatp7}), our results suggest that the orbit of HAT-P-7b
is closer to the polar configuration rather than retrograde as $\lambda$
may imply.

It is worth noting that the suggested discrepancies in $\lambda$ and $v\sin i_\star$ in three data sets (cf. Section \ref{subsec:previous-hatp7}) 
still persist in our analysis. 
For a fair comparison with the A12 result, we repeat the same analyses for the W09 and N09 data only including RVs taken over the same night,
but the values of $\lambda$ and $v\sin i_\star$ do not change significantly.
Since we have used the same model of the RM effect and the same priors from the {\it Kepler} photometry for the three sets of data,
our results confirm that the discrepancy comes from the RV data themselves.
As A12 discussed, such a discrepancy may originate from some physics that is not included in the current model of the RM effect, 
but its origin is beyond the scope of this paper.

As a by-product of the spin--orbit analysis, we have found that HAT-P-7b
has a {small} but non-zero orbital eccentricity, $e=0.005\pm0.001$
(weighted mean of the three data sets), {which is consistent} with
$e=0.0055_{-0.0033}^{+0.007}$ obtained by \citet{2014ApJ...785..126K}.
Our constraint on $e$ comes from the duration and mid-time of the
occultation of HAT-P-7b relative to those of the transit, along with the
constraint on the mean stellar density $\rho_\star$ from
asteroseismology.  This approach is justified by the fact that
$\rho_\star$ from the transit and occultation alone shows a reasonable
agreement with the model stellar density $\rho_{\star,\rm m}$ derived
independently from asteroseismology.  The origin of this non-zero $e$
may {deserve further theoretical consideration} because the tides
are expected to damp $e$ rapidly for such a close-in planet like
HAT-P-7b.

\subsection{Kepler-25\label{subsec:summary_K25}}

For Kepler-25, we obtain
$i_\star=\timeform{65D.4}_{-\timeform{6D.4}}^{+\timeform{10D.6}}$ from
the joint analysis.  This is slightly better than
$i_\star=\timeform{66D.7}^{+\timeform{12D.1}}_{-\timeform{7D.4}}$ from
asteroseismology alone (Figure \ref{fig:is-pdfs-K25}), mainly due to the
prior on $v\sin i_\star$ from spectroscopy.  The constraint on $i_\star$
is better than HAT-P-7 despite the lower signal-to-noise ratio, because
of the greater rotational splitting (see Figure \ref{fig:is-nus-K25}).
This allows us to tightly constrain the spin--orbit angle of Kepler-25c
as $\psi = \timeform{26D.9}_{-\timeform{9D.2}}^{+\timeform{7D.0}}$
(Figure \ref{psi_kep25}). {Our finding is important in two
aspects; 1) this is the first quantitative measurement of $\psi$,
instead of $\lambda$, for multi-planetary systems, except for the
Solar system. 2) Kepler-25 is the first system that exhibits the
significant spin--orbit misalignment among the {multi-transiting systems}
with a main-sequence host star, while it is the second example if we
consider the systems with a red-giant host star, Kepler-56.}

{The spin--orbit misalignment in systems with multiple transiting planets is particularly 
interesting for the following reason.} 
Considering the transit probabilities of multiple planets, planets' orbital planes are likely to
be coplanar in multi-transiting systems, and hence presumably trace
their natal protoplanetary disks.  The spin--orbit misalignment in such
systems, therefore, could be a clue to the processes that tilt a stellar
spin relative to its protoplanetary disk
\citep[e.g.,][]{2010MNRAS.401.1505B, 2011MNRAS.412.2790L,
2012Natur.491..418B}.  

{In this context, the} orbital inclinations of the other two planets (Kepler-25b and
Kepler-25d) relative to that of Kepler-25c would help the interpretation
of the observed misalignment.  They may be constrained from the analysis
of TTVs and Transit Duration Variations (TDVs), along with orbital RVs
to constrain the orbit of the outer non-transiting planet d.  In this
paper, we did not model these phenomena because our main concern is the
determination of the spin--orbit angle.  It should be noted, however,
that the independent information on $\rho_\star$ from asteroseismology
benefits the TTV analysis as well because TTVs are sensitive to the mean
stellar density and orbital eccentricity of the planets
\citep[e.g.,][]{2012Natur.487..449S, 2014ApJ...783...53M}.\\

{
It is also interesting to note that both HAT-P-7 and Kepler-25 are relatively hot stars with $T_{\rm eff}
\gtrsim 6300\,\mathrm{K}$ and in line with the observed trend that the
spin--orbit misalignments are preferentially found around stars with
$T_{\rm eff} > 6250\,\mathrm{K}$ \citep{Winn2010b}.
Although \cite{2012ApJ...758L...6R} suggested that temporal variations of
the stellar rotation due to internal gravity waves could explain this empirical trend,
we found no evidence to support this scenario for the two systems. 
Regarding} {HAT-P-7, we compared
the rotational splitting from Figure \ref{fig:is-pdfs-HATP7} with that
from Q0 to Q2 \citep[results from the study of][]{Oshagh2013}, but
found no evidence of significant variations. Although results using only
Q0 to Q2 have large uncertainties, this may indicate that the rotation
remains constant over time. Moreover, we tightly constrained the
rotation of Kepler-25 and showed that outer layers certainly rotate at
constant velocity. This is incompatible with the scenario suggested by \cite{2012ApJ...758L...6R}.}

\section{Summary 
\label{sec:9}}

The major purpose of the present paper is two-fold. The first is to
develop and describe a detailed methodology of determining the
three-dimensional spin--orbit angle $\psi$ for transiting planetary
systems. The other is to demonstrate the power of the methodology by
applying to the two {specific systems}, HAT-P-7 and
Kepler-25.

The application of asteroseismology to exoplanetary systems is now
becoming popular. It is particularly useful in determining the stellar
inclination $i_\star$ with respect to the line-of-sight. Combined with
the orbital inclination $i_{\rm orb}$ determined for transiting systems,
and with the projected spin-orbit angle $\lambda$ via the spectroscopic
observation of the Rossiter-McLaughlin effect, the joint analysis
{presented in this paper} indeed enables the determination of
$\psi$, rather than $\lambda$. While the observed distribution of
$\lambda$ for more than 70 transiting systems [e.g., figure 7 of
\citet{Xue2014}] already put tight constraints on planetary migration
scenarios, that of $\psi$ is even more useful because it is free from
the projection effect.  As we discussed, HAT-P-7 seems to host a
polar-orbit planet instead of a retrograde one as {naively}
suspected from the observed $\lambda \approx \timeform{180D}$. The
determination of $\psi$ is also important for multi-transiting planetary
systems where all the planets are supposed to share the same orbital
plane; large $\psi$ in such a system indicates that the stellar
obliquity experiences significant tilt with respect to the
protoplanetary disk that would be the orbital plane of the planets. This
turned out to be the case for Kepler-25 as we discussed in the previous
section. 
{While it may be premature to consider the statistics at this point, it is tempting to note that two out of the {six multi-transiting systems with measured spin--orbit angles are shown to be significantly misaligned.
The misaligned cases are Kepler-25c ($\psi = \timeform{26D.9}_{-\timeform{9D.2}}^{+\timeform{7D.0}}$) and Kepler-56 \citep[$i_\star = \timeform{47D} \pm \timeform{6D}$,][]{huber2013}, while the aligned cases are Kepler-30 \citep[$\lambda \lesssim \timeform{10D}$,][]{2012Natur.487..449S}, Kepler-50  and Kepler-65 \citep[$i_\star = \timeform{82D}^{+\timeform{8D}}_{-\timeform{7D}}$ and $i_\star = \timeform{81D}^{+\timeform{9D}}_{-\timeform{16D}}$,][]{Chaplin2013} and Kepler-89d  (KOI-94d) with $\lambda = -\timeform{6D}^{+\timeform{13D}}_{-\timeform{11D}}$ \citep{Hirano2012b} or $-\timeform{11D} \pm \timeform{11D}$ (A13). }
Even if the spin--orbit misalignment is rare, the physical mechanism for its origin is an interesting theoretical question. If it indeed turns out to be fairly common, it will pose a serious challenge to all viable theories of the formation and evolution of multi-planetary systems.}

In addition to the determination of $\psi$, the joint analysis improves
the accuracy and precision of numerous system parameters for a specific
target. In turn, any discrepancy among the separate analyses strongly
points to a certain physical process which needs to be taken into
account in the detailed modeling. This would open a new window for the
exploration of the origin and evolution of planetary systems.

\bigskip 

We are grateful to Simon Albrecht and Josh Winn for providing us with the radial velocity data of Kepler-25.
We thank NASA and the {\it Kepler} team for their revolutionary data.
O.B. is supported by Japan Society for Promotion of Science (JSPS)
Fellowship for Research (No. 25-13316).  K.M. is supported {by JSPS Research Fellowships for Young Scientists  (No. 26-7182) and} 
by the Leading Graduate Course for Frontiers of Mathematical Sciences and Physics.
Y.S. gratefully acknowledges the support from the Grant-in Aid for
Scientific Research by JSPS (No. 24340035).

\bibliographystyle{apj}

\begin{thebibliography}{93}
\expandafter\ifx\csname natexlab\endcsname\relax\def\natexlab#1{#1}\fi

\bibitem[{{Aerts} {et~al.}(2010){Aerts}, {Christensen-Dalsgaard}, \&
  {Kurtz}}]{Aerts2010}
{Aerts}, C., {Christensen-Dalsgaard}, J., \& {Kurtz}, W. 2010,
  Asteroseismology, 1st edn. (Springer Science)

\bibitem[{{Albrecht} {et~al.}(2013){Albrecht}, {Winn}, {Marcy}, {Howard},
  {Isaacson}, \& {Johnson}}]{Albrecht2013}
{Albrecht}, S., {Winn}, J.~N., {Marcy}, G.~W., {Howard}, A.~W., {Isaacson}, H.,
  \& {Johnson}, J.~A. 2013, \apj, 771, 11 (A13)

\bibitem[{{Albrecht} {et~al.}(2012){Albrecht}, {Winn}, {Johnson}, {Howard},
  {Marcy}, {Butler}, {Arriagada}, {Crane}, {Shectman}, {Thompson}, {Hirano},
  {Bakos}, \& {Hartman}}]{2012ApJ...757...18A}
{Albrecht}, S., {et~al.} 2012, \apj, 757, 18 (A12)

\bibitem[{{Appourchaux} {et~al.}(2008){Appourchaux}, {Michel}, {Auvergne},
  {Baglin}, {Toutain}, {Baudin}, {Benomar}, {Chaplin}, {Deheuvels}, {Samadi},
  {Verner}, {Boumier}, {Garc{\'{\i}}a}, {Mosser}, {Hulot}, {Ballot}, {Barban},
  {Elsworth}, {Jim{\'e}nez-Reyes}, {Kjeldsen}, {R{\'e}gulo}, \&
  {Roxburgh}}]{Appourchaux2008}
{Appourchaux}, T., {et~al.} 2008, \aap, 488, 705

\bibitem[{{Appourchaux} {et~al.}(2012){Appourchaux}, {Chaplin},
  {Garc{\'{\i}}a}, {Gruberbauer}, {Verner}, {Antia}, {Benomar}, {Campante},
  {Davies}, {Deheuvels}, {Handberg}, {Hekker}, {Howe}, {R{\'e}gulo},
  {Salabert}, {Bedding}, {White}, {Ballot}, {Mathur}, {Silva Aguirre},
  {Elsworth}, {Basu}, {Gilliland}, {Christensen-Dalsgaard}, {Kjeldsen},
  {Uddin}, {Stumpe}, \& {Barclay}}]{Appourchaux2012}
---. 2012, \aap, 543, A54

\bibitem[{{Asplund} {et~al.}(2009){Asplund}, {Grevesse}, {Sauval}, \&
  {Scott}}]{Asplund2009}
{Asplund}, M., {Grevesse}, N., {Sauval}, A.~J., \& {Scott}, P. 2009, \araa, 47,
  481

\bibitem[{{Baglin} {et~al.}(2006{\natexlab{a}}){Baglin}, {Auvergne}, {Barge},
  {Deleuil}, {Catala}, {Michel}, {Weiss}, \& {COROT Team}}]{Baglin2006a}
{Baglin}, A., {Auvergne}, M., {Barge}, P., {Deleuil}, M., {Catala}, C.,
  {Michel}, E., {Weiss}, W., \& {COROT Team}. 2006{\natexlab{a}}, in ESA
  Special Publication, Vol. 1306, ESA Special Publication, ed. M.~{Fridlund},
  A.~{Baglin}, J.~{Lochard}, \& L.~{Conroy}, 33

\bibitem[{{Baglin} {et~al.}(2006{\natexlab{b}}){Baglin}, {Auvergne},
  {Boisnard}, {Lam-Trong}, {Barge}, {Catala}, {Deleuil}, {Michel}, \&
  {Weiss}}]{Baglin2006b}
{Baglin}, A., {et~al.} 2006{\natexlab{b}}, in COSPAR Meeting, Vol.~36, 36th
  COSPAR Scientific Assembly, 3749

\bibitem[{{Ballot}(2010)}]{ballot2010}
{Ballot}, J. 2010, Astronomische Nachrichten, 331, 933

\bibitem[{{Bate} {et~al.}(2010){Bate}, {Lodato}, \&
  {Pringle}}]{2010MNRAS.401.1505B}
{Bate}, M.~R., {Lodato}, G., \& {Pringle}, J.~E. 2010, \mnras, 401, 1505

\bibitem[{{Batygin}(2012)}]{2012Natur.491..418B}
{Batygin}, K. 2012, \nat, 491, 418

\bibitem[{{Bazot} {et~al.}(2005){Bazot}, {Vauclair}, {Bouchy}, \&
  {Santos}}]{Bazot2005}
{Bazot}, M., {Vauclair}, S., {Bouchy}, F., \& {Santos}, N.~C. 2005, \aap, 440,
  615

\bibitem[{{Benomar} {et~al.}(2009{\natexlab{a}}){Benomar}, {Appourchaux}, \&
  {Baudin}}]{Benomar2009}
{Benomar}, O., {Appourchaux}, T., \& {Baudin}, F. 2009{\natexlab{a}}, \aap,
  506, 15

\bibitem[{{Benomar} {et~al.}(2009{\natexlab{b}}){Benomar}, {Baudin},
  {Campante}, {Chaplin}, {Garc{\'{\i}}a}, {Gaulme}, {Toutain}, {Verner},
  {Appourchaux}, {Ballot}, {Barban}, {Elsworth}, {Mathur}, {Mosser},
  {R{\'e}gulo}, {Roxburgh}, {Auvergne}, {Baglin}, {Catala}, {Michel}, \&
  {Samadi}}]{Benomar2009b}
{Benomar}, O., {et~al.} 2009{\natexlab{b}}, \aap, 507, L13

\bibitem[{{Benomar} {et~al.}(2013){Benomar}, {Bedding}, {Mosser}, {Stello},
  {Belkacem}, {Garcia}, {White}, {Kuehn}, {Deheuvels}, \&
  {Christensen-Dalsgaard}}]{Benomar2013a}
---. 2013, \apj, 767, 158

\bibitem[{{Benomar} {et~al.}(2014){Benomar}, {Belkacem}, {Bedding}, {Stello},
  {Di Mauro}, {Ventura}, {Mosser}, {Goupil}, {Samadi}, \&
  {Garcia}}]{Benomar2014}
---. 2014, \apjl, 781, L29

\bibitem[{{Bevington}(1969)}]{1969drea.book.....B}
{Bevington}, P.~R. 1969, {Data reduction and error analysis for the physical
  sciences}

\bibitem[{{Borucki} {et~al.}(2010){Borucki}, {Koch}, {Basri}, {Batalha},
  {Brown}, {Caldwell}, {Caldwell}, {Christensen-Dalsgaard}, {Cochran},
  {DeVore}, {Dunham}, {Dupree}, {Gautier}, {Geary}, {Gilliland}, {Gould},
  {Howell}, {Jenkins}, {Kondo}, {Latham}, {Marcy}, {Meibom}, {Kjeldsen},
  {Lissauer}, {Monet}, {Morrison}, {Sasselov}, {Tarter}, {Boss}, {Brownlee},
  {Owen}, {Buzasi}, {Charbonneau}, {Doyle}, {Fortney}, {Ford}, {Holman},
  {Seager}, {Steffen}, {Welsh}, {Rowe}, {Anderson}, {Buchhave}, {Ciardi},
  {Walkowicz}, {Sherry}, {Horch}, {Isaacson}, {Everett}, {Fischer}, {Torres},
  {Johnson}, {Endl}, {MacQueen}, {Bryson}, {Dotson}, {Haas}, {Kolodziejczak},
  {Van Cleve}, {Chandrasekaran}, {Twicken}, {Quintana}, {Clarke}, {Allen},
  {Li}, {Wu}, {Tenenbaum}, {Verner}, {Bruhweiler}, {Barnes}, \&
  {Prsa}}]{Borucki2010}
{Borucki}, W.~J., {et~al.} 2010, Science, 327, 977

\bibitem[{{Carter} {et~al.}(2012){Carter}, {Agol}, {Chaplin}, {Basu},
  {Bedding}, {Buchhave}, {Christensen-Dalsgaard}, {Deck}, {Elsworth},
  {Fabrycky}, {Ford}, {Fortney}, {Hale}, {Handberg}, {Hekker}, {Holman},
  {Huber}, {Karoff}, {Kawaler}, {Kjeldsen}, {Lissauer}, {Lopez}, {Lund},
  {Lundkvist}, {Metcalfe}, {Miglio}, {Rogers}, {Stello}, {Borucki}, {Bryson},
  {Christiansen}, {Cochran}, {Geary}, {Gilliland}, {Haas}, {Hall}, {Howard},
  {Jenkins}, {Klaus}, {Koch}, {Latham}, {MacQueen}, {Sasselov}, {Steffen},
  {Twicken}, \& {Winn}}]{Carter2012}
{Carter}, J.~A., {et~al.} 2012, Science, 337, 556

\bibitem[{{Chaplin} {et~al.}(2013){Chaplin}, {Sanchis-Ojeda}, {Campante},
  {Handberg}, {Stello}, {Winn}, {Basu}, {Christensen-Dalsgaard}, {Davies},
  {Metcalfe}, {Buchhave}, {Fischer}, {Bedding}, {Cochran}, {Elsworth},
  {Gilliland}, {Hekker}, {Huber}, {Isaacson}, {Karoff}, {Kawaler}, {Kjeldsen},
  {Latham}, {Lund}, {Lundkvist}, {Marcy}, {Miglio}, {Barclay}, \&
  {Lissauer}}]{Chaplin2013}
{Chaplin}, W.~J., {et~al.} 2013, \apj, 766, 101

\bibitem[{{Christensen-Dalsgaard}(2008)}]{JCD2008b}
{Christensen-Dalsgaard}, J. 2008, \apss, 316, 113

\bibitem[{{Christensen-Dalsgaard} {et~al.}(2010){Christensen-Dalsgaard},
  {Kjeldsen}, {Brown}, {Gilliland}, {Arentoft}, {Frandsen}, {Quirion},
  {Borucki}, {Koch}, \& {Jenkins}}]{JCD2010}
{Christensen-Dalsgaard}, J., {et~al.} 2010, \apjl, 713, L164

\bibitem[{{Claret}(2000)}]{2000A&A...363.1081C}
{Claret}, A. 2000, \aap, 363, 1081

\bibitem[{{Collier Cameron} {et~al.}(2010){Collier Cameron}, {Bruce}, {Miller},
  {Triaud}, \& {Queloz}}]{2010MNRAS.403..151C}
{Collier Cameron}, A., {Bruce}, V.~A., {Miller}, G.~R.~M., {Triaud},
  A.~H.~M.~J., \& {Queloz}, D. 2010, \mnras, 403, 151

\bibitem[{{Dawson} \& {Johnson}(2012)}]{2012ApJ...756..122D}
{Dawson}, R.~I., \& {Johnson}, J.~A. 2012, \apj, 756, 122

\bibitem[{{Esteves} {et~al.}(2013){Esteves}, {De Mooij}, \&
  {Jayawardhana}}]{2013ApJ...772...51E}
{Esteves}, L.~J., {De Mooij}, E.~J.~W., \& {Jayawardhana}, R. 2013, \apj, 772,
  51

\bibitem[{{Fabrycky} \& {Tremaine}(2007)}]{Fabrycky2007}
{Fabrycky}, D., \& {Tremaine}, S. 2007, \apj, 669, 1298

\bibitem[{{Garc{\'{\i}}a} {et~al.}(2011{\natexlab{a}}){Garc{\'{\i}}a},
  {Salabert}, {Ballot}, {Sato}, {Mathur}, \& {Jim{\'e}nez}}]{Garcia2011b}
{Garc{\'{\i}}a}, R.~A., {Salabert}, D., {Ballot}, J., {Sato}, K., {Mathur}, S.,
  \& {Jim{\'e}nez}, A. 2011{\natexlab{a}}, Journal of Physics Conference
  Series, 271, 012049

\bibitem[{{Garc{\'{\i}}a} {et~al.}(2011{\natexlab{b}}){Garc{\'{\i}}a},
  {Hekker}, {Stello}, {Guti{\'e}rrez-Soto}, {Handberg}, {Huber}, {Karoff},
  {Uytterhoeven}, {Appourchaux}, {Chaplin}, {Elsworth}, {Mathur}, {Ballot},
  {Christensen-Dalsgaard}, {Gilliland}, {Houdek}, {Jenkins}, {Kjeldsen},
  {McCauliff}, {Metcalfe}, {Middour}, {Molenda-Zakowicz}, {Monteiro}, {Smith},
  \& {Thompson}}]{Garcia2011}
{Garc{\'{\i}}a}, R.~A., {et~al.} 2011{\natexlab{b}}, \mnras, 414, L6

\bibitem[{{Gizon} \& {Solanki}(2003)}]{Gizon2003}
{Gizon}, L., \& {Solanki}, S.~K. 2003, \apj, 589, 1009

\bibitem[{{Gizon} {et~al.}(2013){Gizon}, {Ballot}, {Michel}, {Stahn},
  {Vauclair}, {Bruntt}, {Quirion}, {Benomar}, {Vauclair}, {Appourchaux},
  {Auvergne}, {Baglin}, {Barban}, {Baudin}, {Bazot}, {Campante}, {Catala},
  {Chaplin}, {Creevey}, {Deheuvels}, {Dolez}, {Elsworth}, {Garcia}, {Gaulme},
  {Mathis}, {Mathur}, {Mosser}, {Regulo}, {Roxburgh}, {Salabert}, {Samadi},
  {Sato}, {Verner}, {Hanasoge}, \& {Sreenivasan}}]{Gizon2013}
{Gizon}, L., {et~al.} 2013, Proceedings of the National Academy of Science,
  110, 13267

\bibitem[{{Grec} {et~al.}(1983){Grec}, {Fossat}, \& {Pomerantz}}]{Grec1983}
{Grec}, G., {Fossat}, E., \& {Pomerantz}, M.~A. 1983, \solphys, 82, 55

\bibitem[{{Guzik} {et~al.}(2014){Guzik}, {Chaplin}, {Handler}, \&
  {Pigulski}}]{2014IAUS..301.....G}
{Guzik}, J.~A., {Chaplin}, W.~J., {Handler}, G., \& {Pigulski}, A., eds. 2014,
  IAU Symposium, Vol. 301, {Precision Asteroseismology}

\bibitem[{Harvey(1985)}]{harvey1985}
Harvey, J. 1985, ESA SP, 235, 199

\bibitem[{{Hirano} {et~al.}(2012{\natexlab{a}}){Hirano}, {Sanchis-Ojeda},
  {Takeda}, {Narita}, {Winn}, {Taruya}, \& {Suto}}]{Hirano2012}
{Hirano}, T., {Sanchis-Ojeda}, R., {Takeda}, Y., {Narita}, N., {Winn}, J.~N.,
  {Taruya}, A., \& {Suto}, Y. 2012{\natexlab{a}}, \apj, 756, 66

\bibitem[{{Hirano} {et~al.}(2010){Hirano}, {Suto}, {Taruya}, {Narita}, {Sato},
  {Johnson}, \& {Winn}}]{Hirano2010}
{Hirano}, T., {Suto}, Y., {Taruya}, A., {Narita}, N., {Sato}, B., {Johnson},
  J.~A., \& {Winn}, J.~N. 2010, \apj, 709, 458

\bibitem[{{Hirano} {et~al.}(2011){Hirano}, {Suto}, {Winn}, {Taruya}, {Narita},
  {Albrecht}, \& {Sato}}]{Hirano2011}
{Hirano}, T., {Suto}, Y., {Winn}, J.~N., {Taruya}, A., {Narita}, N.,
  {Albrecht}, S., \& {Sato}, B. 2011, \apj, 742, 69

\bibitem[{{Hirano} {et~al.}(2012{\natexlab{b}}){Hirano}, {Narita}, {Sato},
  {Takahashi}, {Masuda}, {Takeda}, {Aoki}, {Tamura}, \& {Suto}}]{Hirano2012b}
{Hirano}, T., {et~al.} 2012{\natexlab{b}}, \apjl, 759, L36

\bibitem[{{Huber} {et~al.}(2011){Huber}, {Bedding}, {Stello}, {Hekker},
  {Mathur}, {Mosser}, {Verner}, {Bonanno}, {Buzasi}, {Campante}, {Elsworth},
  {Hale}, {Kallinger}, {Silva Aguirre}, {Chaplin}, {De Ridder},
  {Garc{\'{\i}}a}, {Appourchaux}, {Frandsen}, {Houdek}, {Molenda-{\.Z}akowicz},
  {Monteiro}, {Christensen-Dalsgaard}, {Gilliland}, {Kawaler}, {Kjeldsen},
  {Broomhall}, {Corsaro}, {Salabert}, {Sanderfer}, {Seader}, \&
  {Smith}}]{Huber2011}
{Huber}, D., {et~al.} 2011, \apj, 743, 143

\bibitem[{{Huber} {et~al.}(2013{\natexlab{a}}){Huber}, {Chaplin},
  {Christensen-Dalsgaard}, {Gilliland}, {Kjeldsen}, {Buchhave}, {Fischer},
  {Lissauer}, {Rowe}, {Sanchis-Ojeda}, {Basu}, {Handberg}, {Hekker}, {Howard},
  {Isaacson}, {Karoff}, {Latham}, {Lund}, {Lundkvist}, {Marcy}, {Miglio},
  {Silva Aguirre}, {Stello}, {Arentoft}, {Barclay}, {Bedding}, {Burke},
  {Christiansen}, {Elsworth}, {Haas}, {Kawaler}, {Metcalfe}, {Mullally}, \&
  {Thompson}}]{Huber2013b}
---. 2013{\natexlab{a}}, \apj, 767, 127

\bibitem[{{Huber} {et~al.}(2013{\natexlab{b}}){Huber}, {Carter}, {Barbieri},
  {Miglio}, {Deck}, {Fabrycky}, {Montet}, {Buchhave}, {Chaplin}, {Hekker},
  {Montalb{\'a}n}, {Sanchis-Ojeda}, {Basu}, {Bedding}, {Campante},
  {Christensen-Dalsgaard}, {Elsworth}, {Stello}, {Arentoft}, {Ford},
  {Gilliland}, {Handberg}, {Howard}, {Isaacson}, {Johnson}, {Karoff},
  {Kawaler}, {Kjeldsen}, {Latham}, {Lund}, {Lundkvist}, {Marcy}, {Metcalfe},
  {Silva Aguirre}, \& {Winn}}]{huber2013}
---. 2013{\natexlab{b}}, Science, 342, 331

\bibitem[{{Jackson} {et~al.}(2012){Jackson}, {Lewis}, {Barnes}, {Drake Deming},
  {Showman}, \& {Fortney}}]{2012ApJ...751..112J}
{Jackson}, B.~K., {Lewis}, N.~K., {Barnes}, J.~W., {Drake Deming}, L.,
  {Showman}, A.~P., \& {Fortney}, J.~J. 2012, \apj, 751, 112

\bibitem[{Jeffreys(1961)}]{Jeffreys1961}
Jeffreys, H. 1961, Theory of Probability, 3rd edn. (Oxford, England: Oxford)

\bibitem[{{Kipping}(2014)}]{2014MNRAS.440.2164K}
{Kipping}, D.~M. 2014, \mnras, 440, 2164

\bibitem[{{Kjeldsen} {et~al.}(2008){Kjeldsen}, {Bedding}, \&
  {Christensen-Dalsgaard}}]{Kjeldsen2008}
{Kjeldsen}, H., {Bedding}, T.~R., \& {Christensen-Dalsgaard}, J. 2008, \apjl,
  683, L175

\bibitem[{{Knutson} {et~al.}(2014){Knutson}, {Fulton}, {Montet}, {Kao}, {Ngo},
  {Howard}, {Crepp}, {Hinkley}, {Bakos}, {Batygin}, {Johnson}, {Morton}, \&
  {Muirhead}}]{2014ApJ...785..126K}
{Knutson}, H.~A., {et~al.} 2014, \apj, 785, 126

\bibitem[{{Lai}(2012)}]{Lai2012}
{Lai}, D. 2012, \mnras, 423, 486

\bibitem[{{Lai} {et~al.}(2011){Lai}, {Foucart}, \& {Lin}}]{2011MNRAS.412.2790L}
{Lai}, D., {Foucart}, F., \& {Lin}, D.~N.~C. 2011, \mnras, 412, 2790

\bibitem[{{Lebreton} \& {Montalb{\'a}n}(2009)}]{Lebreton2009}
{Lebreton}, Y., \& {Montalb{\'a}n}, J. 2009, in IAU Symposium, Vol. 258, IAU
  Symposium, ed. {E.~E.~Mamajek, D.~R.~Soderblom, \& R.~F.~G.~Wyse}, 419--430

\bibitem[{{Mandel} \& {Agol}(2002)}]{2002ApJ...580L.171M}
{Mandel}, K., \& {Agol}, E. 2002, \apjl, 580, L171

\bibitem[{{Marcy} {et~al.}(2014){Marcy}, {Isaacson}, {Howard}, {Rowe},
  {Jenkins}, {Bryson}, {Latham}, {Howell}, {Gautier}, {Batalha}, {Rogers},
  {Ciardi}, {Fischer}, {Gilliland}, {Kjeldsen}, {Christensen-Dalsgaard},
  {Huber}, {Chaplin}, {Basu}, {Buchhave}, {Quinn}, {Borucki}, {Koch}, {Hunter},
  {Caldwell}, {Van Cleve}, {Kolbl}, {Weiss}, {Petigura}, {Seager}, {Morton},
  {Johnson}, {Ballard}, {Burke}, {Cochran}, {Endl}, {MacQueen}, {Everett},
  {Lissauer}, {Ford}, {Torres}, {Fressin}, {Brown}, {Steffen}, {Charbonneau},
  {Basri}, {Sasselov}, {Winn}, {Sanchis-Ojeda}, {Christiansen}, {Adams},
  {Henze}, {Dupree}, {Fabrycky}, {Fortney}, {Tarter}, {Holman}, {Tenenbaum},
  {Shporer}, {Lucas}, {Welsh}, {Orosz}, {Bedding}, {Campante}, {Davies},
  {Elsworth}, {Handberg}, {Hekker}, {Karoff}, {Kawaler}, {Lund}, {Lundkvist},
  {Metcalfe}, {Miglio}, {Silva Aguirre}, {Stello}, {White}, {Boss}, {Devore},
  {Gould}, {Prsa}, {Agol}, {Barclay}, {Coughlin}, {Brugamyer}, {Mullally},
  {Quintana}, {Still}, {Thompson}, {Morrison}, {Twicken}, {D{\'e}sert},
  {Carter}, {Crepp}, {H{\'e}brard}, {Santerne}, {Moutou}, {Sobeck}, {Hudgins},
  {Haas}, {Robertson}, {Lillo-Box}, \& {Barrado}}]{2014ApJS..210...20M}
{Marcy}, G.~W., {et~al.} 2014, \apjs, 210, 20

\bibitem[{{Masuda}(2014)}]{2014ApJ...783...53M}
{Masuda}, K. 2014, \apj, 783, 53

\bibitem[{{Masuda} {et~al.}(2013){Masuda}, {Hirano}, {Taruya}, {Nagasawa}, \&
  {Suto}}]{Masuda2013}
{Masuda}, K., {Hirano}, T., {Taruya}, A., {Nagasawa}, M., \& {Suto}, Y. 2013,
  \apj, 778, 185

\bibitem[{{McLaughlin}(1924)}]{M1924}
{McLaughlin}, D.~B. 1924, \apj, 60, 22

\bibitem[{{Metcalfe} {et~al.}(2012){Metcalfe}, {Chaplin}, {Appourchaux},
  {Garc{\'{\i}}a}, {Basu}, {Brand{\~a}o}, {Creevey}, {Deheuvels}, {Do{\v g}an},
  {Eggenberger}, {Karoff}, {Miglio}, {Stello}, {Y{\i}ld{\i}z}, {{\c C}elik},
  {Antia}, {Benomar}, {Howe}, {R{\'e}gulo}, {Salabert}, {Stahn}, {Bedding},
  {Davies}, {Elsworth}, {Gizon}, {Hekker}, {Mathur}, {Mosser}, {Bryson},
  {Still}, {Christensen-Dalsgaard}, {Gilliland}, {Kawaler}, {Kjeldsen},
  {Ibrahim}, {Klaus}, \& {Li}}]{Metcalfe2012}
{Metcalfe}, T.~S., {et~al.} 2012, \apjl, 748, L10

\bibitem[{{Monteiro} {et~al.}(1994){Monteiro}, {Christensen-Dalsgaard}, \&
  {Thompson}}]{Monteiro1994}
{Monteiro}, M.~J.~P.~F.~G., {Christensen-Dalsgaard}, J., \& {Thompson}, M.~J.
  1994, \aap, 283, 247

\bibitem[{{Morris} {et~al.}(2013){Morris}, {Mandell}, \&
  {Deming}}]{2013ApJ...764L..22M}
{Morris}, B.~M., {Mandell}, A.~M., \& {Deming}, D. 2013, \apjl, 764, L22

\bibitem[{{Nagasawa} \& {Ida}(2011)}]{Nagasawa2011}
{Nagasawa}, M., \& {Ida}, S. 2011, \apj, 742, 72

\bibitem[{{Nagasawa} {et~al.}(2008){Nagasawa}, {Ida}, \&
  {Bessho}}]{Nagasawa2008}
{Nagasawa}, M., {Ida}, S., \& {Bessho}, T. 2008, \apj, 678, 498

\bibitem[{{Narita} {et~al.}(2009){Narita}, {Sato}, {Hirano}, \&
  {Tamura}}]{Narita2009}
{Narita}, N., {Sato}, B., {Hirano}, T., \& {Tamura}, M. 2009, \pasj, 61, L35
  (N09)

\bibitem[{{Narita} {et~al.}(2012){Narita}, {Takahashi}, {Kuzuhara}, {Hirano},
  {Suenaga}, {Kandori}, {Kudo}, {Sato}, {Suzuki}, {Ida}, {Nagasawa}, {Abe},
  {Brandner}, {Brandt}, {Carson}, {Egner}, {Feldt}, {Goto}, {Grady}, {Guyon},
  {Hashimoto}, {Hayano}, {Hayashi}, {Hayashi}, {Henning}, {Hodapp}, {Ishii},
  {Iye}, {Janson}, {Knapp}, {Kusakabe}, {Kwon}, {Matsuo}, {Mayama}, {McElwain},
  {Miyama}, {Morino}, {Moro-Martin}, {Nishimura}, {Pyo}, {Serabyn}, {Suto},
  {Takami}, {Takato}, {Terada}, {Thalmann}, {Tomono}, {Turner}, {Watanabe},
  {Wisniewski}, {Yamada}, {Takami}, {Usuda}, \& {Tamura}}]{2012PASJ...64L...7N}
{Narita}, N., {et~al.} 2012, \pasj, 64, L7

\bibitem[{{Nelder} \& {Mead}(1965)}]{Simplex}
{Nelder}, J.~A., \& {Mead}, R. 1965, The Computer Journal, 7, 308

\bibitem[{{Ohta} {et~al.}(2005){Ohta}, {Taruya}, \& {Suto}}]{Ohta2005}
{Ohta}, Y., {Taruya}, A., \& {Suto}, Y. 2005, \apj, 622, 1118

\bibitem[{{Ohta} {et~al.}(2009){Ohta}, {Taruya}, \&
  {Suto}}]{2009ApJ...690....1O}
---. 2009, \apj, 690, 1

\bibitem[{{Oshagh} {et~al.}(2013){Oshagh}, {Grigahc{\`e}ne}, {Benomar},
  {Dupret}, {Monteiro}, {Scuflaire}, \& {Santos}}]{Oshagh2013}
{Oshagh}, M., {Grigahc{\`e}ne}, A., {Benomar}, O., {Dupret}, M.-A., {Monteiro},
  M.~J.~P.~F.~G., {Scuflaire}, R., \& {Santos}, N.~C. 2013, in Astrophysics and
  Space Science Proceedings, Vol.~31, Stellar Pulsations: Impact of New
  Instrumentation and New Insights, ed. J.~C. {Su{\'a}rez}, R.~{Garrido}, L.~A.
  {Balona}, \& J.~{Christensen-Dalsgaard}, 227

\bibitem[{{P{\'a}l} {et~al.}(2008){P{\'a}l}, {Bakos}, {Torres}, {Noyes},
  {Latham}, {Kov{\'a}cs}, {Marcy}, {Fischer}, {Butler}, {Sasselov}, {Sip{\H
  o}cz}, {Esquerdo}, {Kov{\'a}cs}, {Stefanik}, {L{\'a}z{\'a}r}, {Papp}, \&
  {S{\'a}ri}}]{Pal2008}
{P{\'a}l}, A., {et~al.} 2008, \apj, 680, 1450 (P08)

\bibitem[{{Paxton} {et~al.}(2011){Paxton}, {Bildsten}, {Dotter}, {Herwig},
  {Lesaffre}, \& {Timmes}}]{paxton2011}
{Paxton}, B., {Bildsten}, L., {Dotter}, A., {Herwig}, F., {Lesaffre}, P., \&
  {Timmes}, F. 2011, \apjs, 192, 3

\bibitem[{{Paxton} {et~al.}(2013){Paxton}, {Cantiello}, {Arras}, {Bildsten},
  {Brown}, {Dotter}, {Mankovich}, {Montgomery}, {Stello}, {Timmes}, \&
  {Townsend}}]{paxton2013}
{Paxton}, B., {et~al.} 2013, \apjs, 208, 4

\bibitem[{{Pont} {et~al.}(2006){Pont}, {Zucker}, \&
  {Queloz}}]{2006MNRAS.373..231P}
{Pont}, F., {Zucker}, S., \& {Queloz}, D. 2006, \mnras, 373, 231

\bibitem[{{Queloz} {et~al.}(2000){Queloz}, {Eggenberger}, {Mayor}, {Perrier},
  {Beuzit}, {Naef}, {Sivan}, \& {Udry}}]{Queloz2000}
{Queloz}, D., {Eggenberger}, A., {Mayor}, M., {Perrier}, C., {Beuzit}, J.~L.,
  {Naef}, D., {Sivan}, J.~P., \& {Udry}, S. 2000, \aap, 359, L13

\bibitem[{{Reese} {et~al.}(2006){Reese}, {Ligni{\`e}res}, \&
  {Rieutord}}]{Reese2006}
{Reese}, D., {Ligni{\`e}res}, F., \& {Rieutord}, M. 2006, \aap, 455, 621

\bibitem[{{Rogers} {et~al.}(2012){Rogers}, {Lin}, \&
  {Lau}}]{2012ApJ...758L...6R}
{Rogers}, T.~M., {Lin}, D.~N.~C., \& {Lau}, H.~H.~B. 2012, \apjl, 758, L6

\bibitem[{{Rossiter}(1924)}]{R1924}
{Rossiter}, R.~A. 1924, \apj, 60, 15

\bibitem[{{Roxburgh} \& {Vorontsov}(2003)}]{Roxburgh2003}
{Roxburgh}, I.~W., \& {Vorontsov}, S.~V. 2003, \aap, 411, 215

\bibitem[{{Sanchis-Ojeda} {et~al.}(2012){Sanchis-Ojeda}, {Fabrycky}, {Winn},
  {Barclay}, {Clarke}, {Ford}, {Fortney}, {Geary}, {Holman}, {Howard},
  {Jenkins}, {Koch}, {Lissauer}, {Marcy}, {Mullally}, {Ragozzine}, {Seader},
  {Still}, \& {Thompson}}]{2012Natur.487..449S}
{Sanchis-Ojeda}, R., {et~al.} 2012, \nat, 487, 449

\bibitem[{{Schlaufman}(2010)}]{2010ApJ...719..602S}
{Schlaufman}, K.~C. 2010, \apj, 719, 602

\bibitem[{{Seager} \& {Mall{\'e}n-Ornelas}(2003)}]{2003ApJ...585.1038S}
{Seager}, S., \& {Mall{\'e}n-Ornelas}, G. 2003, \apj, 585, 1038

\bibitem[{{Shibahashi} \& {Lynas-Gray}(2013)}]{2013ASPC..479.....S}
{Shibahashi}, H., \& {Lynas-Gray}, A.~E., eds. 2013, Astronomical Society of
  the Pacific Conference Series, Vol. 479, {Progress in Physics of the Sun and
  Stars}

\bibitem[{{Shibahashi} {et~al.}(2012){Shibahashi}, {Takata}, \&
  {Lynas-Gray}}]{2012ASPC..462.....S}
{Shibahashi}, H., {Takata}, M., \& {Lynas-Gray}, A.~E., eds. 2012, Astronomical
  Society of the Pacific Conference Series, Vol. 462, {Progress in
  Solar/Stellar Physics with Helio- and Asteroseismology}

\bibitem[{{Shporer} \& {Brown}(2011)}]{2011ApJ...733...30S}
{Shporer}, A., \& {Brown}, T. 2011, \apj, 733, 30

\bibitem[{{Steffen} {et~al.}(2012){Steffen}, {Fabrycky}, {Ford}, {Carter},
  {D{\'e}sert}, {Fressin}, {Holman}, {Lissauer}, {Moorhead}, {Rowe},
  {Ragozzine}, {Welsh}, {Batalha}, {Borucki}, {Buchhave}, {Bryson}, {Caldwell},
  {Charbonneau}, {Ciardi}, {Cochran}, {Endl}, {Everett}, {Gautier},
  {Gilliland}, {Girouard}, {Jenkins}, {Horch}, {Howell}, {Isaacson}, {Klaus},
  {Koch}, {Latham}, {Li}, {Lucas}, {MacQueen}, {Marcy}, {McCauliff}, {Middour},
  {Morris}, {Mullally}, {Quinn}, {Quintana}, {Shporer}, {Still}, {Tenenbaum},
  {Thompson}, {Twicken}, \& {Van Cleve}}]{2012MNRAS.421.2342S}
{Steffen}, J.~H., {et~al.} 2012, \mnras, 421, 2342

\bibitem[{Unno {et~al.}(1989)Unno, Osaki, Ando, Saio, \& Shibahashi}]{Unno1989}
Unno, W., Osaki, Y., Ando, H., Saio, H., \& Shibahashi, H. 1989, {Nonradial
  Oscillations of Stars} (University Tokyo Press)

\bibitem[{{Van Eylen} {et~al.}(2012){Van Eylen}, {Kjeldsen},
  {Christensen-Dalsgaard}, \& {Aerts}}]{VanEylen2012}
{Van Eylen}, V., {Kjeldsen}, H., {Christensen-Dalsgaard}, J., \& {Aerts}, C.
  2012, Astronomische Nachrichten, 333, 1088

\bibitem[{{Van Eylen} {et~al.}(2013){Van Eylen}, {Lindholm Nielsen}, {Hinrup},
  {Tingley}, \& {Kjeldsen}}]{2013ApJ...774L..19V}
{Van Eylen}, V., {Lindholm Nielsen}, M., {Hinrup}, B., {Tingley}, B., \&
  {Kjeldsen}, H. 2013, \apjl, 774, L19

\bibitem[{{Van Eylen} {et~al.}(2014){Van Eylen}, {Lund}, {Silva Aguirre},
  {Arentoft}, {Kjeldsen}, {Albrecht}, {Chaplin}, {Isaacson}, {Pedersen},
  {Jessen-Hansen}, {Tingley}, {Christensen-Dalsgaard}, {Aerts}, {Campante}, \&
  {Bryson}}]{2014ApJ...782...14V}
{Van Eylen}, V., {et~al.} 2014, \apj, 782, 14

\bibitem[{{Vorontsov}(1988)}]{Vorontsov1988}
{Vorontsov}, S.~V. 1988, in IAU Symposium, Vol. 123, Advances in Helio- and
  Asteroseismology, ed. J.~{Christensen-Dalsgaard} \& S.~{Frandsen}, 151

\bibitem[{{Walker} {et~al.}(2003){Walker}, {Matthews}, {Kuschnig}, {Johnson},
  {Rucinski}, {Pazder}, {Burley}, {Walker}, {Skaret}, {Zee}, {Grocott},
  {Carroll}, {Sinclair}, {Sturgeon}, \& {Harron}}]{2003PASP..115.1023W}
{Walker}, G., {et~al.} 2003, \pasp, 115, 1023

\bibitem[{{White} {et~al.}(2012){White}, {Bedding}, {Gruberbauer}, {Benomar},
  {Stello}, {Appourchaux}, {Chaplin}, {Christensen-Dalsgaard}, {Elsworth},
  {Garc{\'{\i}}a}, {Hekker}, {Huber}, {Kjeldsen}, {Mosser}, {Kinemuchi},
  {Mullally}, \& {Still}}]{White2012}
{White}, T.~R., {et~al.} 2012, \apjl, 751, L36

\bibitem[{{Winn}(2011)}]{2011exop.book...55W}
{Winn}, J.~N. 2011, in Exoplanets, ed. S.~{Seager} (Tucson, AZ: University of
  Arizona Press), 55--77

\bibitem[{{Winn} {et~al.}(2010){Winn}, {Fabrycky}, {Albrecht}, \&
  {Johnson}}]{Winn2010b}
{Winn}, J.~N., {Fabrycky}, D., {Albrecht}, S., \& {Johnson}, J.~A. 2010, \apjl,
  718, L145

\bibitem[{{Winn} {et~al.}(2009){Winn}, {Johnson}, {Albrecht}, {Howard},
  {Marcy}, {Crossfield}, \& {Holman}}]{Winn2009}
{Winn}, J.~N., {Johnson}, J.~A., {Albrecht}, S., {Howard}, A.~W., {Marcy},
  G.~W., {Crossfield}, I.~J., \& {Holman}, M.~J. 2009, \apjl, 703, L99 (W09)

\bibitem[{{Winn} {et~al.}(2005){Winn}, {Noyes}, {Holman}, {Charbonneau},
  {Ohta}, {Taruya}, {Suto}, {Narita}, {Turner}, {Johnson}, {Marcy}, {Butler},
  \& {Vogt}}]{Winn2005}
{Winn}, J.~N., {et~al.} 2005, \apj, 631, 1215

\bibitem[{{Xue} {et~al.}(2014){Xue}, {Suto}, {Taruya}, {Hirano}, {Fujii}, \&
  {Masuda}}]{Xue2014}
{Xue}, Y., {Suto}, Y., {Taruya}, A., {Hirano}, T., {Fujii}, Y., \& {Masuda}, K.
  2014, \apj, 784, 66

\end{thebibliography}

\end{document}